\documentclass[sigconf]{acmart}
\AtBeginDocument{%
  }

\usepackage{xspace}
\usepackage{xcolor}
\definecolor{bizchat-orange}{HTML}{F88545}

\usepackage{float} 

\newcommand{\prototype}{Prototype\xspace}
\newcommand{\pittsburgh}{Pittsburgh\xspace}

\newcommand{\prototypepgh}{Prototype PGH\xspace}

\newcommand{\change}[1]{\textcolor{black}{#1}}
\newcommand{\sysname}{\textsc{BizChat}\xspace}

\newcommand{\feature}[1]{\textbf{#1}}

\usepackage{tabularx}
\usepackage{array} 
\usepackage{cleveref} 

\copyrightyear{2026}
\acmYear{2026}
\setcopyright{cc}
\setcctype{by}
\acmConference[CHI '26]{Proceedings of the 2026 CHI Conference on Human Factors in Computing Systems}{April 13--17, 2026}{Barcelona, Spain}
\acmBooktitle{Proceedings of the 2026 CHI Conference on Human Factors in Computing Systems (CHI '26), April 13--17, 2026, Barcelona, Spain}
\acmDOI{10.1145/3772318.3791654}
\acmISBN{979-8-4007-2278-3/2026/04}

\begin{document}

\title{Towards Designing for Resilience: Community-Centered Deployment of an AI Business Planning Tool in a Small Business Center}



\begin{abstract}
Entrepreneurs in resource-constrained communities often lack time and support to translate ideas into actionable business plans. 
While generative AI promises assistance, most systems assume high digital literacy and overlook community infrastructures that shape adoption.
We report on the community-centered design and deployment of BizChat, an AI-powered business planning tool, introduced across four workshops at a feminist makerspace in Pittsburgh.
Through log data (N=30) and interviews (N=10), we examine how entrepreneurs build resilience through collective AI literacy development—encompassing adoption, adaptation, and refusal of AI. 
Our findings reveal that while BizChat lowered barriers to accessing capital by translating ideas into "business language," this ease raised questions about whether instant AI outputs undermine sensemaking essential to planning. 
We show how peer support helped entrepreneurs navigate this tension. We contribute design implications, including productive friction, communal scaffolds, and co-optability, for strengthening resilience amid technological change.
\end{abstract}

\ccsdesc[500]{Human-centered computing~Human computer interaction (HCI)}
\keywords{Small Business, Entrepreneurship, Business Planning, Generative AI, Resilience}

\author{Quentin Romero Lauro}
\affiliation{%
  \institution{University of Pittsburgh}
  \city{Pittsburgh}
  \state{PA}
  \country{USA}
}

\author{Aakash Gautam}
\affiliation{%
  \institution{University of Pittsburgh}
  \city{Pittsburgh}
  \state{PA}
  \country{USA}
}

\author{Yasmine Kotturi}
\affiliation{%
  \institution{Univ of Maryland, Baltimore County}
  \city{Baltimore}
  \state{MD}
  \country{USA}
}

\vspace{-2mm}
\begin{teaserfigure}
\label{sec:teaser}
\begin{minipage}[t]{\columnwidth}
    \centering
    \includegraphics[width=0.99\linewidth]{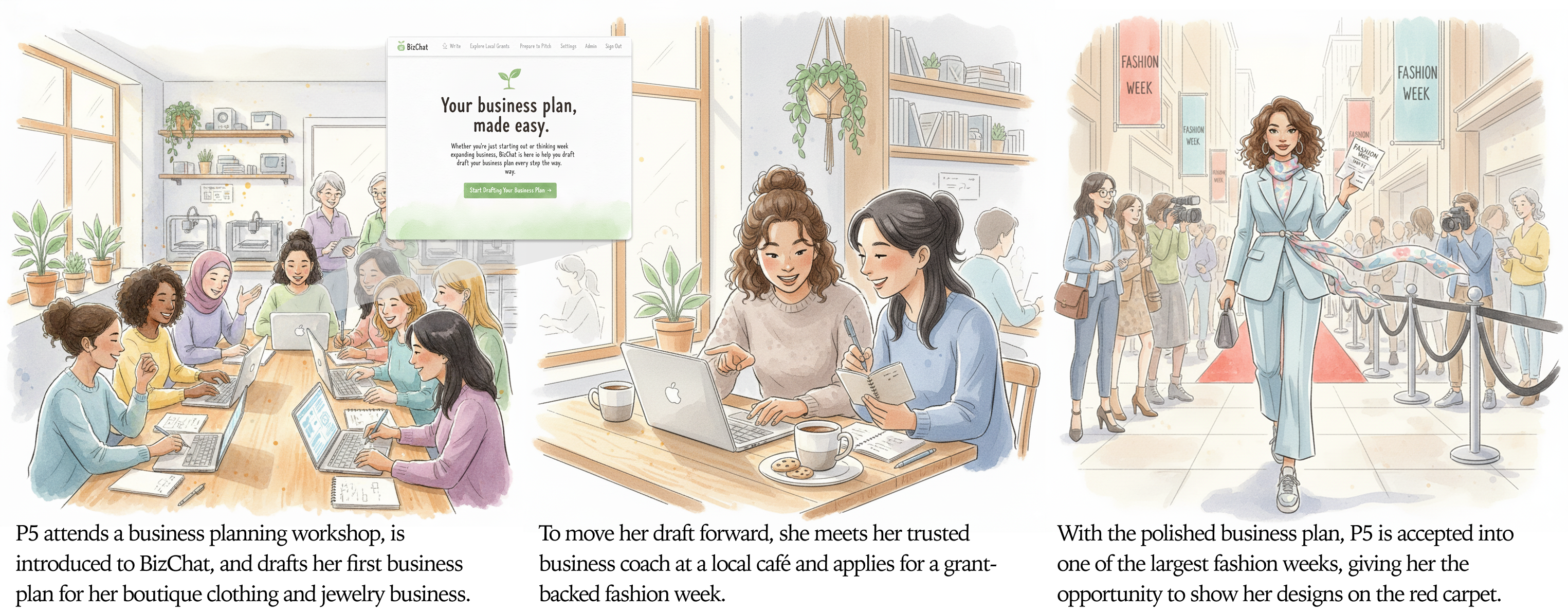}
    \Description[Three-panel watercolor comic of P5's journey]{
      Three-panel watercolor comic showing P5's journey. 
      In the first panel, P5 attends a business planning workshop at a local makerspace, where she is introduced to BizChat and drafts her initial business plan.
      In the second panel, P5 meets with a business coach at a local café to refine and strengthen her plan.
      In the third panel, P5 walks confidently down a red carpet at Fashion Week, holding her acceptance letter, symbolizing that she successfully used the polished plan to showcase her designs.
    }
    \vspace{-1.9mm}
    \caption{We introduced \sysname, an AI-powered business planning tool for small business owners, and deployed it through co-designed, community-centered workshops at a feminist makerspace in Pittsburgh. This figure shows P5, a custom jewelry and apparel designer, using \sysname to create her first business plan and gain entry to a grant-backed fashion week program.}
    \label{fig:teaser}
\end{minipage}
\end{teaserfigure}
\maketitle

\section{Introduction}
Small business owners are uniquely positioned to benefit from generative AI, particularly for automating tedious back-office tasks where time and resource constraints often hinder growth.  
Prior work suggests that large language models (LLMs) can improve entrepreneurs' bottom line by streamlining workflows, soliciting and implementing feedback, and unlocking time for long-term planning~\cite{otis2023uneven}.  
These systems can produce high-quality drafts for written content, support decision-making, and leverage broad knowledge bases to adapt to generalist tasks.  

However, integrating generative AI into entrepreneurial workflows is far from straightforward.  
While commercial systems, such as ChatGPT or Gemini, present themselves as simple to use~\cite{openai_chatgpt_overview}, their successful adoption requires a range of "hidden" operational skills \cite{jackson2002life}.  
Entrepreneurs must navigate browser literacy, password management, cloud storage, and other foundational practices that extend well beyond prompt engineering~\cite{kotturi2024deconstructing}. 
Even when entrepreneurs successfully generate AI outputs, a gap remains in \change{translating generic AI-generated text into contextualized, business-specific content that reflects their actual operations, customers, and goals~\cite{kotturi2024deconstructing}.}

These barriers compound for entrepreneurs in resource-constrained settings, where limited access to social capital, infrastructure, and time for digital upskilling creates widening disparities in AI adoption and outcomes ~\cite{dillahunt2018entrepreneurship,kotturi2022tech,hui2020community,hui2018making}.  
For many small business owners, accessing capital requires business plans---formal documents that articulate goals and strategies for growth~\cite{sba_business_plan}---but writing these plans demands both business planning expertise and comfort with ``the language of business" that funders expect.
This double barrier of needing to both understand what a business plan should contain and possess the writing fluency to articulate it leaves many entrepreneurs unable to compete for funding, even when their business ideas are sound.
While users with higher technical fluency can leverage generative AI to accelerate such tasks, others face widening disparities in use and outcome~\cite{gusto_2024, pewresearch_2023}.  
Over-reliance on generative AI for tasks ill-suited to automation can further exacerbate these divides, reinforcing ``rich-get-richer" dynamics in technology adoption~\cite{otis2023uneven}.  

\change{In this work, we examine how small business owners navigate these compounding challenges through two concepts. 
First, we draw on resilience---patterns of positive adaptation in the context of adversity~\cite{riley2005resilience}---as a collective capacity that emerges from shared resources, mutual support, and local infrastructures~\cite{romero2024exploring}. We operationalize resilience as the shift from uncertainty and hesitation in the face of technological change (``unknown unknowns'') to collectively arriving at possibilities of action (``known unknowns'')~\cite{weick1995sensemaking,termeer2013organizational}.
Second, we use AI literacy to refer to both operational competence with AI systems and the ability to adopt, adapt, or refuse AI tools based on one's context \cite{bhargava2015beyond, long2020ai}.
These concepts are closely intertwined, as AI literacy enables resilience by transforming opaque, threatening technology into something entrepreneurs can evaluate strategically, and resilience emerges when communities collectively develop that literacy rather than navigating technological change alone.
Building on this, we argue that for entrepreneurs in 2025, business resilience and AI literacy are inseparable: the pressure to compete for capital, maintain professional credibility, and stay current with evolving tools means that navigating AI is itself a core business challenge. 
}

We partnered with \prototypepgh, a feminist makerspace and entrepreneurial hub in \pittsburgh, Pennsylvania dedicated to racial and gender equity in technology.  
Building on formative research~\cite{kotturi2024deconstructing}, we co-developed and deployed \sysname\footnote{\textsc{BizChat} can be freely used at: \url{http://bizchat-io.vercel.app/}. 
We named our tool BizChat before Microsoft Copilot Business Chat was released—since the tool is increasingly established within the community of entrepreneurs we serve, we decided not to rename.}, an LLM-powered web application that supports product- and service-based entrepreneurs in writing business plans.
Unlike commercial offerings such as Microsoft 365 Copilot, which rely on users' ability to craft sophisticated prompts~\cite{msf365copilot}, \sysname scaffolds both business planning by centering the entrepreneur's existing knowledge, and adapting the technology to the entrepreneur's context, rather than requiring the entrepreneur to adapt to the technology.  

\change{\sysname's design operationalizes the two concepts—resilience and AI literacy—through three corresponding design goals. }
First, we adopt a low-floor-high-ceiling~\cite{resnick2008falling} approach to aid users with limited digital skills or no prior knowledge of generative AI while offering advanced features for entrepreneurs experienced with tools like ChatGPT, ensuring that the tool supports AI literacy development across the full spectrum from adoption through adaptation to refusal.  
To facilitate skill-building for time-constrained entrepreneurs~\cite{avle2019additional}, our second design consideration is just-in-time learning~\cite{hug2005micro, shail2019using}, which scaffolds AI literacy incrementally as entrepreneurs work toward their business goals rather than requiring upfront training.  
Finally, we explicitly contextualize the introduction of generative AI within the users' existing knowledge and goals (i.e., their business), rather than highlighting the novel technology~\cite{bar2007mobile, gautam2020crafting}, supporting resilience by grounding AI adoption in entrepreneurs' concrete expertise and immediate needs.

From January through August 2025, we integrated \sysname into a four-session workshop series at \prototype.  
\change{Workshops were framed as ``Resilience and Uncertainty Planning'' sessions focused on business planning, with AI introduced as one tool among many for drafting plans. This framing was deliberate: by positioning business planning---not AI---as the focal point, we created space for entrepreneurs to engage with technology on their own terms while surfacing questions about adoption, trust, and control. 
In total, 30 entrepreneurs used BizChat, including workshop participants (N=21) and organic users who adopted through word-of-mouth (N=9). 
We collected system log data and conducted 10 semi-structured follow-up interviews to understand how entrepreneurs navigated AI-assisted business planning within existing support infrastructures.}

Rather than evaluating \sysname in isolation, we investigate how its design and deployment interact with existing practices of support, learning, and adaptation, and how participants navigated the combined challenges of business planning, AI adoption, and capital access.
From this perspective, we ask the following research questions:
\begin{itemize}
    \item RQ1. How should AI systems be designed to lower existing barriers of adoption for entrepreneurs in resource-constrained contexts?
    \item RQ2. How can existing community infrastructures mediate the development of AI literacy—including adoption, adaptation, and refusal of AI tools—for entrepreneurship? 
    \change{\item RQ3. In what ways does resilience emerge when AI tools are deployed within existing communities of entrepreneurial support?}
\end{itemize}

\change{
By designing for low-floor-high-ceiling accessibility (e.g., voice-to-text for entrepreneurs with limited typing proficiency), just-in-time learning (e.g., scaffolded onboarding questions), and contextualized AI introduction (e.g., extracting information from entrepreneurs' existing websites), \sysname aimed to lower barriers to adoption [RQ1].
Our findings reveal that while \sysname reduced barriers to accessing capital by scaffolding the drafting process and ``leveling the language of business,'' this very ease introduced tensions between relying on instant outputs versus engaging in the deeper sensemaking and reflection essential to effective planning [RQ1].
We show how peer support helped participants overcome technology gaps—such as contextualizing generic AI text for their specific customers, judging when outputs were accurate versus plausible-sounding, and translating business jargon into their authentic voice—by collectively developing strategies for editing, refining, and assessing output quality [RQ2].
Further, while \sysname seeded peer-led infrastructures for business planning and AI adoption, participants debated whether AI tools should be introduced by trusted community members or external experts, revealing tensions around credibility, expertise, and the appropriate pace of technology adoption [RQ2].
Participants demonstrated resilience against technological exclusion and capital barriers by collectively transforming their initial fear, uncertainty, and confusion about AI---an ``unknown unknown"---into concrete strategies for adopting, adapting, and refusing AI [RQ3].}

Taken together, this paper contributes: (1) empirical insights from the log data and community-based deployment of \sysname with 30 entrepreneurs; (2) a methodological framing of resilience in technology adoption, embedding it into workshops and accountability interviews; and (3) practical design and deployment implications for introducing AI systems in resource-constrained communities, including the value of productive friction, communal scaffolds for AI literacy, and designing for co-optability.
\section{Related Work}
\subsection{The Hustle Economy Meets AI}
The rapid rise of the ``hustle'' economy---characterized by growing numbers of freelance, solo‐entrepreneurial ventures and side‐hustle culture---reflects shifting work norms where individuals increasingly pursue flexible, short-term, and independent income streams across diverse sectors~\cite{borchers2025sidehustle}. 
Today, nearly half of U.S. small businesses with paid staff operate at an extremely modest scale, with 49\% (or 2.9 million employer firms) employing just one to four workers, showing that much of the small business landscape resembles solopreneurship or micro-enterprise rather than larger employer firms~\cite{leppert2024smallbusiness}.
Technology has accelerated this shift by reducing startup costs, expanding customer reach through e-commerce and mobile platforms, and lowering barriers to entry, widely touted as a ``democratizing force'' for innovation, economic mobility, and social change~\cite{vonhippel2006democratizing}.
While positive narratives of ``hustle culture'' cast entrepreneurial pursuits as aspirational pathways to autonomy and self-realization \cite{Cook2021RiseOfTikTokEntrepreneurs}, HCI and business researchers caution that this framing obscures the economic precarity underlying many ventures, where individuals pursue entrepreneurship less out of choice and more out of necessity~\cite{larsson2019independent, mueller2023necessity}.
Scholars describe this ``necessity-driven entrepreneurship'' as self-employment undertaken in response to limited job opportunities, systemic employment barriers, or unstable economic conditions ---  a pathway for survival rather than aspiration~\cite{hui2018making, mueller2023necessity}.

Against this backdrop, AI raises pressing questions for HCI scholars: will these tools expand opportunities for necessity-driven entrepreneurs, or further entrench divides in access and capacity?
On the one hand, a growing ecosystem of AI tools promises to help small-scale business owners tackle their endless to-do lists through automation and outsourcing (e.g., \textit{Storyteq} for automated ads creation, \textit{Clay} for automated leads generation, business planning platforms like \textit{LivePlan} and \textit{Upmetrics}), indicating promise towards democratizing entrepreneurship among those with internet access.
But in practice, successful adoption requires what scholars call a "hidden curriculum" of digital work \cite{jackson2002life} --- a ``laundry list of operational skills'' entrepreneurs must acquire in order to effectively integrate AI into small business, including basic browser literacy, file-type conversions, password management, and clipboard know-how, among others~\cite{kotturi2024deconstructing}. 
But even after acquiring these accessory skills, entrepreneurs from lean-economies~\cite{dillahunt2018entrepreneurship} face deeper obstacles: resource constraints and historical infrastructural barriers to technology adoption~\cite{hui2020community, avle2019additional}, ultimately perpetuating a ``rich gets richer'' dynamic in AI adoption~\cite{otis2023uneven}.
Addressing such divides requires moving beyond individual solutions to foreground community and structural barriers~\cite{hui2020community}. 
We extend this scholarship by deploying a system addressing these tensions and uneven conditions across design, development, and adoption.

\subsection{\change{Resilience Through Collective Capacity Building}}

Amid rapid technological change, people must adapt their lifestyles, workflows, and everyday practices.
Resilience is widely recognized as a critical capacity for navigating these shifts \cite{Li02092025, masten2001ordinary, riley2005resilience}.
Scholars in psychology and human development define resilience as ``patterns of positive adaptation in the context of past or present adversity, one class of phenomena observed in human lives'' \cite{masten2001ordinary}.
In practice, scholars have detailed various strategies people use to exercise resilience, such as developing new routines to develop emotional stability in the face of loss or trauma~\cite{Bonanno2004LossTrauma}, coping by re-framing challenges as manageable amid chronic stress~\cite{Rutter2012ResilienceDynamic}, or by cultivating self-regulation and persistence to maintain academic performance amid economic hardship~\cite{Luthar1991VulnerabilityResilience}.
However, resilience is not an individual capacity alone \cite{Ungar2013ResilienceContext, Southwick2014ResilienceDefinitions, Patel2017CommunityResilience, Norris2007CommunityResilience}. %
It is also embedded in and functions as a collective capacity. 
\change{Resilience is related to but distinct from social support and collective learning. 
Social support provides resources; collective learning builds skills, but resilience describes how communities transform uncertainty and threat into actionable knowledge and sustained practice \cite{masten2001ordinary, riley2005resilience}.
This distinction matters here, where entrepreneurs face not just skills or resource gaps, but widespread uncertainty about engaging with rapidly evolving, potentially threatening technology.
Resilience captures the adaptive capacity to navigate this uncertainty --- turning ``unknown unknowns'' into ``known unknowns'' that can be named, discussed, and acted on collectively~\cite{weick1995sensemaking,termeer2013organizational}.}

\change{HCI scholarship demonstrates how this transformation can occur in practice.} 
Scholars have shown how people draw on social and material resources within their communities to adapt when facing adversity \cite{vyas2017everyday}. 
For example, scholarship on assets-based design advocates identifying and building on community assets, such as collective knowledge, shared resources, and peer support, as a way to enable communities to realize goals that matter to them (e.g., \cite{wong2021reflections, gautam2020crafting, pei2019we}).
\change{Collective capacity building can scaffold resilience by allowing people to pool resources and knowledge \cite{vyas2017everyday, hui2020community, kotturi2022tech}. 
In entrepreneurial contexts, for instance, \citet{hui2020community} describe ``Community Collectives,'' where small groups of necessity-driven entrepreneurs support one another in overcoming barriers to digital platform use through low-tech coordination and regular gatherings.
Similarly, \citet{kotturi2022tech} examine a ``Tech Help Desk'' model in which a small business support program provided long-term, one-on-one technical assistance.
Through collaboratively addressing the ``long tail'' of everyday computing challenges, entrepreneurs solved immediate problems, and also developed strategic skills (e.g., goal setting) and operational skills (e.g., implementation) for adapting to technological change ~\cite{kotturi2022tech}. 
Other work highlights how community organizations act as trusted intermediaries that broker technical knowledge, translate platform norms, and provide ongoing, place-based support \cite{dillahunt2014fostering, harrington2019deconstructing}.
The scholarship highlights the often hidden, varied, but critical non-technological infrastructures that are necessary in supporting communities, such as entrepreneurs in lean economies \cite{dillahunt2018entrepreneurship, DillahuntEtAl_CHI22_VillageMentoring}. }

While prior scholarship shows how resilience can be scaffolded in response to economic and digital inequities, AI introduces a new layer of fragility.
Recently, \citet{glassman2024ai} propose AI-resilient interfaces designed to help users notice, judge, and recover from AI-generated errors.
These practices are elements of building resilience and can be leveraged to facilitate collective capacity building. For example, when noticing and recovering from errors are facilitated through group-based sensemaking and leveraging domain knowledge that participants already have, it can build collective capacity and strengthen resilience. 
Building on these insights, we position technological capacities within the community and explore how we can support resilience building through a network of support---where our design, research, and deployment are aligned to scaffold community interactions and collective engagement. 

\change{
An equally important part of resilience in this setting is the capacity for refusal. 
Refusal refers to the choice not to use a technology, or to use it selectively, due to misalignment with one’s goals, values, or constraints \cite{satchell2009beyond, selwyn2006digital, baumer2014refusing, gautam2024reconfiguring}. 
For instance, \citet{kotturi2024deconstructing} demonstrate entrepreneurs' hesitation around using AI for grant writing or marketing copy due to fears of inaccuracy or legal ambiguity. 
Designing for resilience, then insists on designing for informed refusal where users are supported to evaluate when and how AI fits their needs.
}

\change{
Building on this scholarship, we frame resilience in our work as the community’s capacity to adopt, adapt, and refuse AI. 
We operationalize this framing through \sysname, a business planning system and associated workshop designed to embed AI use within community contexts and support the collective capacities entrepreneurs need to sustain their ventures amid an unstable technological landscape.
}



\subsection{Community-Based Participatory Design in AI and Entrepreneurship}

Scholars in HCI have long emphasized that design processes risk reproducing existing social inequities if they fail to center the perspectives of those most affected by technological systems~\cite{dillahunt2014fostering, harrington2019deconstructing}.
Community-Based Participatory Research (CBPR)~\cite{Israel1998Review}---a methodological approach aligned with participatory design~\cite{muller2002participatory} (PD)---seeks to equitably involve community members and researchers across all stages of a project.
Within HCI, CBPR has been increasingly taken up to ensure that communities most affected by technology play a role in shaping the design, development, and deployment, and through it, foster an equitable design process (e.g.,\cite{kotturi2022tech, hui2020community, dillahunt2014fostering, liang2021embracing, gautam2022empowering}).

Within entrepreneurial contexts, CPBR's community-centric orientation aligns closely with HCI research that shows how trust, peer network, non-technological efforts, and shared visions are essential in supporting and sustaining entrepreneurs and small businesses \cite{dillahunt2018entrepreneurship, hui2020community, kotturi2024sustaining}.
Community spaces such as makerspaces and entrepreneurial hubs~\cite{Kay2023MakerspacesEffectiveLearning} scaffold digital engagement~\cite{kotturi2022tech, hui2023community} and enable practices like help-seeking and peer learning ~\cite{KuhnGalloway2015PeerNetworking, kotturi2024peerdea} and adaptation to technological change \cite{kotturi2024deconstructing}. 
HCI scholarship draws attention to the value of community-based research in understanding the resources available to the community, the power dynamics that are at play, and the entangled complexities where designed technologies are situated \cite{wong2021reflections, gautam2020crafting, cooper2022systematic}. 

Aligned with this, scholarship describes a growing ``participatory turn'' in AI design, calling for approaches that embed communities most affected by AI systems into the processes of design, development, and governance in ways that extend beyond token consultation toward meaningful influence and shared agency~\cite{delgado2023participatory}.
However, this turn is not without its limitations and critiques. 
Scholars warn that participation in AI design with clarity on why AI is used, and what power participants have in shaping it, can risk cooption and cause harm \cite{birhane2022power, gautam2024reconfiguring}. 
Similarly, other scholars highlight the need to build reciprocity, lest the participation become a one-off ``blue sky ideation'' that privileges unconstrained novelty over the on-the-ground reality that those most affected by technology face \cite{harrington2019deconstructing,parker2025participatory, pierre2021getting}.
These challenges are further amplified by the vast design space and speculative promise associated with emerging AI technologies~\cite{delgado2023participatory, Morris2023DesignSpaceGenerativeModels}

In this work, we take up CBPR in the design and deployment of an AI-embedded business-planning system within a feminist makerspace in Pittsburgh, Pennsylvania.
Our approach builds on a prior four-part workshop series on generative AI and entrepreneurship at another local small business hub~\cite{kotturi2024deconstructing}.  Guided by lessons from these workshops and CBRP's call for building reciprocity, we scoped our design intentionally on a constrained, practical application---business planning---that offers direct, tangible value for local entrepreneurs.
Moreover, we rely on our existing relationship with the makerspace to deploy the technology and examine how its adoption can be scaffolded through the community setting. 
Next, we describe our system design informed by our prior workshop and grounded in entrepreneurs’ needs, the deployment context, and our broader commitment to collective resilience.

\section{\sysname}

\subsection{Business Planning as Context for Resilience Building}
\label{sec:context-for-resilience-building}
We chose business planning as the focus for our system because it provides a practical context for resilience building.
Business plans are essential, yet often overlooked documents, required for entrepreneurs to apply to grants, loans, and small business accelerators.
Beyond accessing capital, business plans are tools for reflection, requiring entrepreneurs to examine goals, strategies, and operations against their expertise.
Contrasting with traditional business planning approaches (e.g., SCORE~\cite{score}), which emphasize static templates, we take an iterative learning-oriented approach to business planning.
Last, we saw that current AI-business planning tools (e.g., LivePlan~\cite{liveplan}, Upmetrics~\cite{upmetrics}, IdeaBuddy~\cite{ideabuddy}) prioritize automated drafting, rather than accessible interfaces and community connection required for resilience building in practice.
Building on these foundations of what makes a business plan, we designed \sysname to instantiate community-centered design principles in the context of AI-supported business planning.

\begin{figure*}
  \centering
  \includegraphics[width=0.95\textwidth]{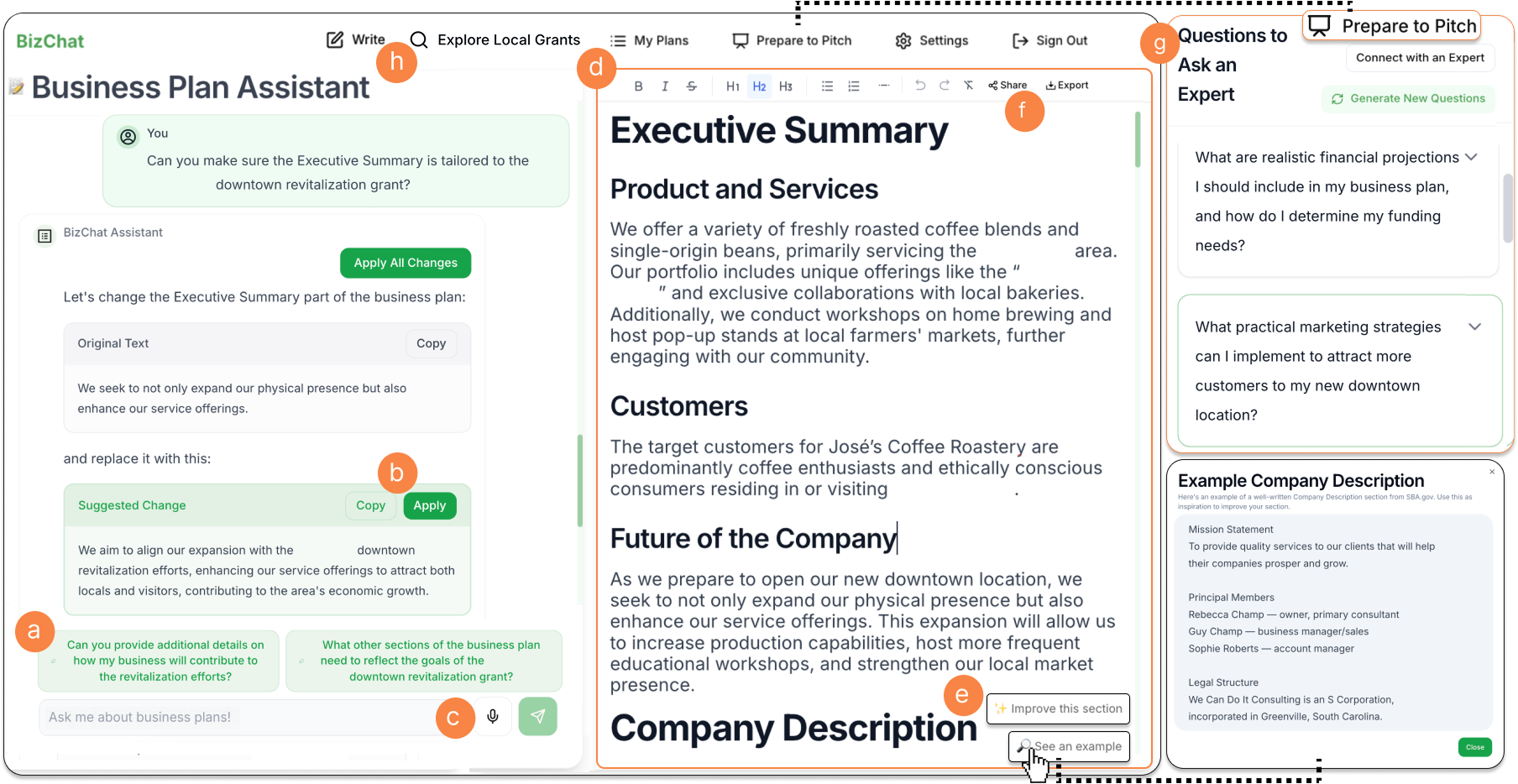}
  \caption{BizChat supports small business owners to draft and refine business plans. Interface legend. (a) Prompt suggestions for the next turn. (b) One-click \emph{Apply} to insert suggested changes into the plan. (c) Voice-to-text for dictation. (d) Rich-text editor for direct manipulation. (e) Inline generation/exemplars to draft or revise selected text. (f) \emph{Export} to a formatted document. (g) \emph{Prepare to Pitch}: expert-style questions and an option to connect with an expert.}
  \Description{This is a screenshot of the BizChat interface's main page. On the left, there is a chat and on the right there is a rich text editor. BizChat is a system that scaffolds drafting, iterating, and getting feedback on a business plan.}
  \label{fig:bizchat-overview}
\end{figure*}

\subsection{Formative Study and Design Goals}
\subsubsection{Formative Study}
The design of \sysname builds directly on a four-part community workshop series and interview with local entrepreneurs focused on integrating (or supporting the refusal to integrate) generative AI into small business workflows~\cite{kotturi2024deconstructing}.
In this workshop series, which focused on practical applications of AI in small business contexts, authors surfaced the long tail of accessory skills (e.g., browser literacy, password management, file literacy, typing skills, clipboard management) that entrepreneurs must acquire in order to effectively use generative AI.
In this section, we detail the findings of this formative study and related work from HCI and learning sciences that led to our three design goals for \sysname: (1) support diverse levels of digital and business expertise through low-floors and high-ceilings, (2) embed just-in-time learning opportunities that connect AI use directly to business outcomes, and (3) contextualize AI introduction within entrepreneurs’ existing knowledge and goals to reduce anxiety and build confidence.

\subsubsection{Low Floors, High Ceilings}
Previous workshops revealed that entrepreneurs came with highly diverse technical and business backgrounds. 
Some struggled with accessory skills such as typing or file management, while others were already experimenting simultaneously with many AI tools, such as ChatGPT, DALL-E, and Midjourney, among others. 
Thus, \sysname's first design consideration is to build ``low-floors and high-ceilings''---a pedagogical term which refers to designing interventions that are accessible to beginners while providing opportunities for advanced learners to engage in deeper exploration~\cite{papert2020mindstorms}.
For small business owners, this means creating a system accessible to users with limited digital skills, while remaining extensible for those already familiar with tools like ChatGPT.
As small business owners have diverse levels of typing and writing proficiency---from two-finger typing to touch typing---we scaffold the editing process so that \feature{Voice-to-Text (\ref{fig:bizchat-overview}c)} is available in every step: describing the business, dictating edits, and chatting with \sysname.
For users comfortable with typing, \sysname's \feature{Rich-text Editor (\ref{fig:bizchat-overview}d)} enables more control through direct manipulation of their business plan.
As prompting remains a barrier for non-expert users of LLMs~\cite{zamfirescu2023johnny}, \sysname creates two \feature{Prompt Suggestions (\ref{fig:bizchat-overview}a)} for every conversation turn---one prompt suggestion focuses on the current topic (exploitation); the other focuses on a new topic (exploration)~\cite{Pirolli2007InformationForaging}.
When \sysname provides suggestions, the user can one-click \feature{Apply (\ref{fig:bizchat-overview}b)} suggestions from the chat to the business plan in the rich-text editor.
For more advanced users, \sysname enables in-line text generation \feature{(\ref{fig:bizchat-overview}e)}, where users can specify criteria to generate text and view exemplars.
Further, the \feature{Rich-text Editor (\ref{fig:bizchat-overview}d)} allows users without word-processing skills to specify text styles and \feature{Export (\ref{fig:bizchat-overview}f)} to a standard template, without the additional barriers of complex document formatting.

\subsubsection{Just-in-Time Learning} 
Prior design workshops revealed that for entrepreneurs from resource-constrained backgrounds, the value of general-purpose AI is often unclear~\cite{kotturi2024deconstructing}.
Further, participants emphasized that, given their busy schedules, they could not step away from their businesses for AI literacy training or generic AI tutorials.
Thus, \sysname's second design consideration is just-in-time learning~\cite{hug2005micro,shail2019using}---in situ micro-learning opportunities that support entrepreneurial education.
\sysname aims to develop skills in three areas: understanding effective business plans, building reflection and help-seeking skills, and effectively using generative AI.
To do so, \sysname's onboarding flow includes an \feature{Informational Video} reviewing key motivations for business plan creation and detailing their key components.
\sysname then scaffolds users to iterate on each section of their business plan, suggesting relevant \feature{Examples (\ref{fig:bizchat-overview}e)} from U.S. Small Business Administration\footnote{\url{https://www.sba.gov/business-guide/plan-your-business/write-your-business-plan}}.
In addition, \sysname provides users with relevant suggested questions as tooltips in the editor for the \feature{Business Plan Assistant (\ref{fig:bizchat-overview}e)}.
Last, to build help-seeking skills and facilitate self-reflection, \sysname's \feature{Prepare to Pitch (\ref{fig:bizchat-overview}g)} provides users a list of questions to ask an expert tailored to their business plan and goal, which are a helpful preparatory step when approaching business coaches for critical feedback.

\subsubsection{Contextualized Introduction to Technology}
Prior design workshops also revealed how entrepreneurs express anxiety about falling behind or ``missing out'' on new technological trends, which compounds hesitation to engage with unfamiliar tools~\cite{kotturi2024deconstructing}. 
To mitigate such anxieties, introducing technology in the context of users’ existing knowledge and goals---rather than focusing on the novel technology---can be a helpful strategy~\cite{bar2007mobile, gautam2020crafting}.
Therefore, \sysname's third design consideration is to contextualize users' interactions with generative AI within their area of expertise---their business~\cite{hidi2006four}.
For instance, at each conversation turn, the \feature{Business Plan Assistant (\ref{fig:bizchat-overview}b)} suggests changes to the business plan in accordance with the user’s \feature{Business Plan Goals}, which are explicitly set during onboarding. 
Further, unlike AI systems that present outputs as a completed document, \sysname positions business plans as an evolving document used as a tool for planning and growth. 
To this end, \sysname seeks to facilitate deeper engagement with a broader community of support (e.g., expert business coaches, an essential part of an entrepreneurial ecosystem~\cite{hui2020community}).
\sysname offers users to \feature{Connect with an Expert (\ref{fig:bizchat-overview}g)} and aims to decrease reputational risks by providing entrepreneurs a list of \feature{Questions to Ask an Expert (\ref{fig:bizchat-overview}g)} about their business plan.
In doing so, \sysname acknowledges that, even with the latest technology, social support among minority entrepreneurs is essential~\cite{kotturi2024sustaining, kotturi2024peerdea, kotturi2022tech, hui2023community}.

\begin{figure*}
  \centering
  \includegraphics[width=1\textwidth]{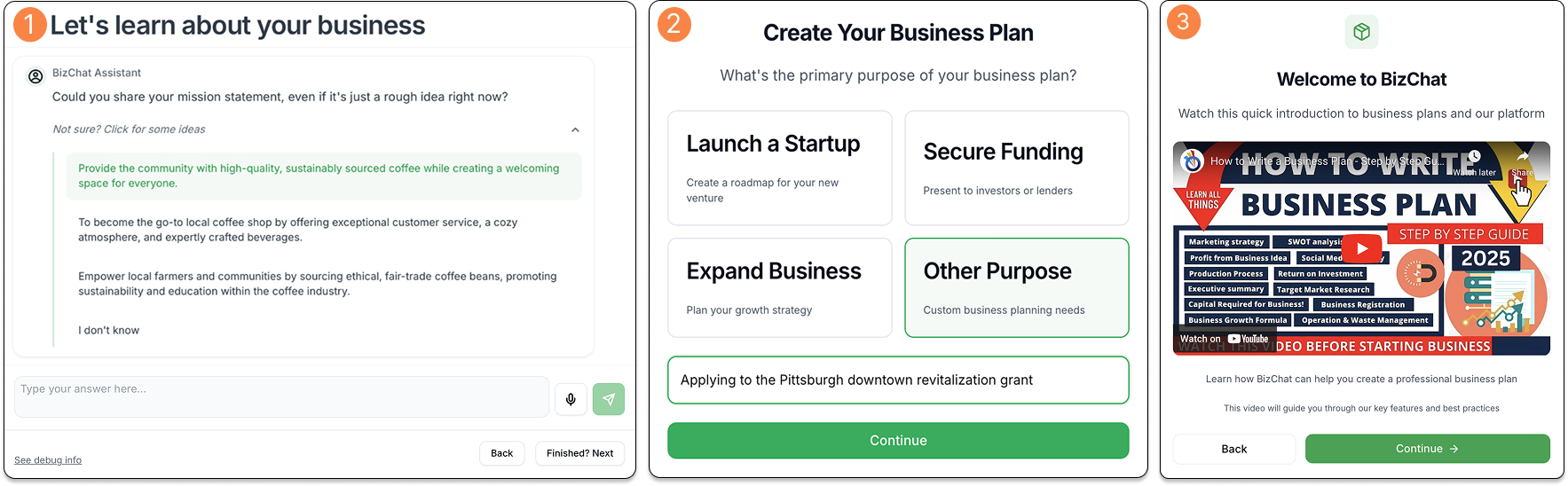}
  \caption{BizChat scaffolds the onboarding process for business plan creation. To \textbf{(1) collect business context}, users can enter their website or chat with the assistant about their idea. Next, users \textbf{(2) specify the purpose} of their business plan, such as starting a new venture, applying for funding, or expanding an existing business. Users can optionally \textbf{(3) watch an informational video} that introduces business plans and the BizChat workflow. Finally, \sysname automatically generates a first draft of the business plan based on the provided information.}
  \Description{This figure shows the BizChat onboarding flow. Step (1) allows entering a website or chatting to share business information. Step (3) prompts the user to select the purpose of their plan. Step (2) offers an optional video introducing business plans. Step (d) shows BizChat generating a first draft of the business plan.}
  \label{fig:bizchat-onboarding}
\end{figure*}

\subsection{\sysname User Scenario}
To illustrate how \sysname can be used to draft, iterate, and take next steps with a business plan, let us follow Tina, an aspiring entrepreneur with a boutique clothing company.
Although she runs her business from her mobile phone, primarily selling through Facebook Marketplace and Etsy, like many other solo-preneurs~\cite{stripe_stripe_nodate}, she finds the maintenance of this constellation of digital tools is tedious and takes her away from core business activities
As a result, keeping pace with rapidly advancing technologies, like ChatGPT or Gemini, is challenging and overwhelming~\cite{kotturi2024deconstructing}. Moreover, Tina finds it unclear how to extract value from these tools long-term.

\subsubsection{The need for a business plan}
After three years of consecutive four-digit revenue proving the viability of her small business, Tina wants to expand her business into a brick-and-mortar store in her city's downtown.
While attending a community pop-up market for local artists, her solo-preneur friend informs her about a recent funding opportunity to revitalize her city's downtown storefronts.
However, to apply for the funding and subsidized shopfront that she needs to scale up her business, 
Tina, like many other small business owners on the cusp of growth, needs a business plan.
While reviewing the requirements for the grant application, Tina finds \sysname on the resource page.

\subsubsection{Drafting a Business Plan with \sysname}
After logging in, Tina is directed to \sysname's onboarding, which asks her to enter her website, but also gives her the option to proceed by just chatting if she does not have one prepared. Tina enters the URL of the website her niece vibe-coded for her.
\sysname then extracts relevant information from the website.
But, since the website is incomplete without essential details for a business plan (e.g., product information, legal structure), \sysname detects the missing details and asks Tina for the information it is missing.
\sysname asks: ``What's your vision statement?''
Not lacking a vision, but unclear on how to precisely articulate it, Tina clicks ``Not sure? Click for some ideas.'' to view \sysname's response suggestions (\Cref{fig:bizchat-onboarding}.1).
After \sysname has all of the information needed for a first draft, it generates a first draft of Tina's business plan.
While waiting, Tina watches a just-in-time learning module about the sections of a business plan.
While reviewing the business plan \sysname generates, she notices an error in the Executive Summary.
Because her two-finger typing is less efficient, she uses \sysname's voice-to-text to edit the plan in the rich-text editor.
\sysname suggests changes, and after reviewing, Tina clicks ``Apply'' and watches \sysname make the changes to her business plan, highlighting exactly where it is making the change.
\sysname's prompt suggestions \feature{(\Cref{fig:bizchat-overview}a)}, then suggest making some changes to the Market Analysis section.
To verify the changes suggested, Tina compares with an example Market Analysis small business plan from SBA.gov~\cite{SBA2025WriteBusinessPlan} in the \sysname interface.

With Tina's newfound understanding of what an effective market analysis section looks like, she presses ``Apply'' and applies the changes.
While Tina has no interest in complex file-formatting for business plans, she clicks ``Export'' to get a pre-formatted business plan in a standard format to share with others.
With a draft business plan in hand, Tina wants to prepare for the grant application. 
She navigates to the ``Prepare to Pitch'' tab in the \sysname interface.
Previously apprehensive to share her plan with her mentors, she reviews \sysname's custom ``Questions to Ask an Expert,'' and feels more prepared to share her business plan before applying for grants.

\subsubsection{Overcoming technology gaps with her local network}
\begin{figure}[t]
  \centering
  \includegraphics[width=\columnwidth]{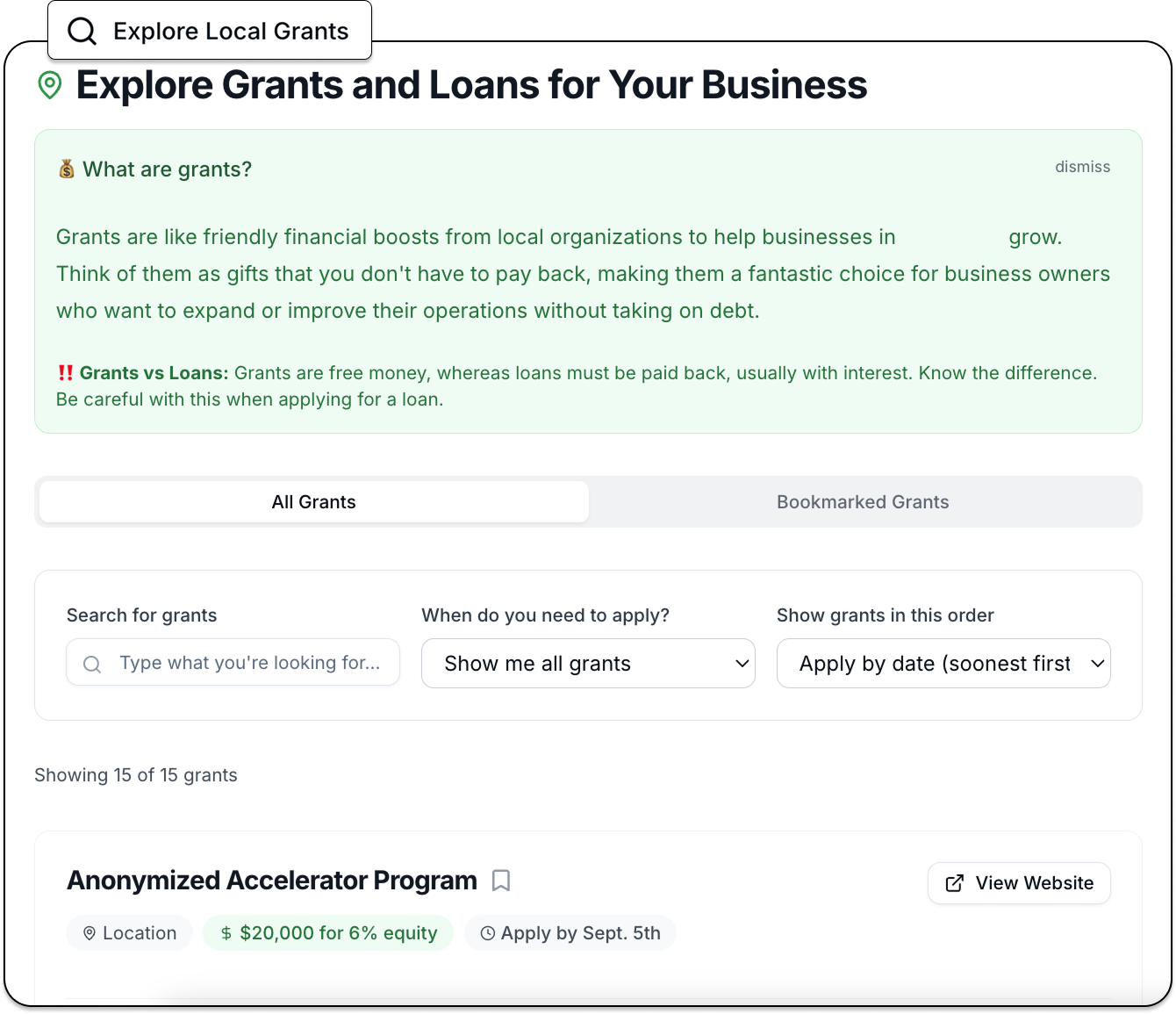}
  \caption{\sysname's \emph{Explore Local Grants} tab provides a curated list of local grants and loans maintained by the research team. Users can search, filter by deadlines, and view detailed information about each opportunity, including funding amount, equity requirements, and application dates. This feature helps entrepreneurs discover and apply for relevant opportunities without leaving the platform.}
  \Description{Screenshot of the Explore Local Grants interface in BizChat. The interface includes a description of what grants are, a comparison of grants versus loans, a search bar, filters for deadlines, and a list of grants with funding amount, equity, and application deadlines.}
  \label{fig:bizchat-local-grants}
\end{figure}

With a draft in hand, Tina shared her business plan with her close entrepreneurial friends.
While she felt clear about her product and vision, she was unsure how to build the financial projections required for the grant.
Tina's mentor walked her through revenue and expense estimates, filling in gaps Tina had struggled to overcome alone with \sysname.
Meeting the requirements for the grant application, Tina is able to apply to the revitalization grant and explore more opportunities in \sysname's \feature{Explore Local Grants} tab (See \Cref{fig:bizchat-local-grants}).
Encouraged by this experience, Tina shares \sysname with other trusted entrepreneurs in her network.

\subsection{Implementation}
\sysname is implemented as a React application built on Next.js, with Firebase handling authentication, data storage, and telemetry. 
\sysname is built on top of TipTap for its rich text editor and is deployed with Vercel.
At the core of \sysname's business plan generation is a structured business plan JSON that is progressively filled out during onboarding. 
Each field is initialized with a \texttt{NOT\_FOUND} placeholder and is dynamically updated as the user provides information. 
The onboarding assistant, powered by \texttt{GPT-4 Turbo Preview}, asks only about missing fields and adapts questions accordingly. 
At each conversation turn, an observer LLM powered by \texttt{GPT-4} extracts business information from users' answers to onboarding questions and updates the business plan JSON. 

When a website is provided, \sysname crawls within-domain pages and uses \texttt{GPT-4} with function-calling to extract structured information. 
The onboarding assistant dynamically adjusts what questions to ask based on what information is missing from the business plan JSON.
If no website is provided, the chat begins immediately. 
To ensure enough context is provided to each business plan prompt section, minimal completion requires a name, mission, target market, at least one product with a value proposition, and a basic marketing and sales strategy.

Once the structured business plan JSON meets completion criteria, \sysname makes asynchronous calls to \texttt{gpt-4-turbo} to generate each section, with only relevant structured information inserted into the prompt. 
This approach maintains interactive latencies (around 8–15 seconds per section) instead of waiting for a single long response. Each business plan section's prompt is few-shot with examples provided from SBA.gov exemplar business~\cite{SBA2025WriteBusinessPlan}. 
\sysname's business plan assistant is powered by \textbf{GPT-4o} and is prompted to generate \texttt{<suggestion>} tags, which render in-chat components that allow users to directly apply suggestions to their business plan by clicking a button.
We intentionally did not implement automated evaluation metrics. 
Instead, to reinforce accountability with our community partner, the generated plans were shared with community partners for a sanity check, focusing on practical utility in local contexts.

\section{Methods}

\subsection{Community Site}
Our research was conducted in partnership with \prototypepgh (or ``\prototype'' for short), a feminist makerspace located in Pittsburgh, Pennsylvania.
Founded in 2016, \prototype's mission is to foster gender and racial equity in creative entrepreneurship by ``providing affordable access to high tech tools and equipment, offering workshops that prioritize the experiences of marginalized communities, and cultivating a professional support network.''
\prototype's ethos is that ``everything is a prototype,'' emphasizing feedback and iteration as essential to early-stage creative and entrepreneurial practice.

\prototype functions as both a makerspace and an entrepreneurial hub, providing shared equipment (e.g., laser cutters, 3D printers, and sewing machines), skill-building workshops, and a supportive peer network. 
Its programming is intentionally designed to serve underrepresented entrepreneurs, particularly women and gender minorities, and people of color, who often face compounded barriers to accessing resources, technical skills, and professional support.

Our collaboration with \prototype was built on longstanding relationships between the members of the research team and the community site.
One author has been a dues-paying member and collaborator for several years (2017-2025), participating in both the makerspace and its business incubator programs. 
This positionality (see \autoref{sec:positionality}) facilitated trust with \prototype leadership and participants, and allowed us to ground the design and deployment of \sysname within a community already invested in inclusive, iterative approaches to entrepreneurship.

Participants were recruited through the authors' existing local networks, developed over years of community-based research in entrepreneurship and technology. 
Recruitment materials were circulated via personal and professional networks on platforms including LinkedIn, X (formerly Twitter), and via \prototype's existing communications channels (i.e., Instagram, newsletter), extending reach to its broader membership and entrepreneurial community.

\subsection{Four Workshops (N=21)}





We developed and deployed \sysname through a series of four workshops hosted in collaboration with \prototype between January and August 2025. 
The first workshop (January 2025) was integrated into \prototype's annual incubator program as part of a session on financial planning.
The two-hour workshop was structured in two parts: during the first hour, facilitators introduced key financial documents and strategies for small business owners. 
One author was present throughout onboarding to provide live support, helping participants navigate technical issues and ensuring they could begin drafting their business plans with the tool.

\change{This framing was a deliberate methodological choice.
Rather than centering AI in our outreach and workshop titles, we positioned business planning as the focal point.
We did so to create an accessible entry point for entrepreneurs who might feel anxious about AI-focused framing, an issue highlighted in prior work~\cite{kotturi2024deconstructing}, to ground engagement in participants' existing business expertise, and to observe how entrepreneurs engage with AI on their own terms within a task they already needed to complete. 
We reflect on the tensions this choice introduced---including concerns about transparency raised by one community partner---in Section 6.2.}

Following positive reception, we were invited to return for three additional standalone workshops dedicated to \sysname and business planning. 
We hosted three workshops that took place on Saturdays in June, July, and August 2025. 
Each of these sessions followed a consistent structure under the title ``Write Your Business Plan: Resilience and Uncertainty Planning Workshop.'' 
The workshops began with content on the basic structure of a business plan and on the role of business planning as a resilience practice, positioning business plans as living, evolving documents that can adjust based on the needs of the business.
We explicitly framed business plans as drafts requiring ongoing iteration, emphasizing that participants could create different versions for different goals (e.g., grant applications vs. loan proposals).
\change{This framing was intentional. By positioning the business plan, and not the AI technology, as the workshop's focal point, we sought to ground participants' engagement in their existing business expertise rather than foregrounding technical concerns about AI.}

After introductory content, participants were invited to use \sysname to create a draft business plan. 
As in the first workshop, a research team member was present to answer questions and help workshop participants overcome technical difficulties while using the tool.
\change{Rather than beginning with abstract explanations of LLM limitations (e.g., hallucinations, training data biases), we relied on participants' expertise to surface issues as they arose during use.
For instance, when participants noticed AI-generated content that did not align with their business knowledge---such as incorrect product descriptions or generic market analysis that did not reflect their actual customer base—these observations prompted group-wide discussions about AI accuracy, the importance of human oversight, and strategies for evaluating output quality.
This organic approach to discussing AI limitations allowed participants to identify issues relevant to their specific contexts rather than requiring them to navigate abstract technical terminology before understanding its practical implications.}

Following their use of \sysname, participants engaged in reflection activities with paper-based planning tools that scaffolded resilience strategies such as if–then planning (shown to strengthen adaptive responses under uncertainty \cite{gollwitzer1999implementation}), goal setting, reflecting on challenges, and writing down next steps for accountability.
These post-use activities were designed to complement \sysname by encouraging participants to identify concrete next steps beyond the AI-generated draft (e.g., connecting with mentors, researching funding opportunities, refining financial projections with expert support).

\begin{table}[t]
    \centering
    \newcolumntype{L}[1]{>{\raggedright\arraybackslash}p{#1}}
    \centering
    \begin{tabular}{@{}L{0.1\linewidth}L{0.86\linewidth}@{}}
        \toprule
        \textbf{PID} & \textbf{Business Description} \\
        \midrule
        P1 & Designer and creative director; freelance design business owner \\
        P2 & Co-founder, makerspace for neurodivergent vocational training \\
        P3 & Co-founder, makerspace for neurodivergent vocational training \\
        P4 & Gift baskets and party planning business \\
        P5 & Custom jewelry and apparel designer \\
        P6 & Psychic and medium \\
        P7 & Nuisance control business; youth mentor at Job Corps \\
        P8 & Founder, consulting firm advising micro-businesses and nonprofits \\
        P9 & Director of a local makerspace; writer and novelist \\
        P10 & Licensed counselor \& missionary \\
        \bottomrule
    \end{tabular}
    \Description[Table showing participants' businesses]{This table presents descriptions of the nine participants’ businesses. Businesses span creative industries (design, jewelry, apparel), community organizations (makerspace for neurodivergent vocational training, consulting for micro-businesses and nonprofits), service sectors (party planning, nuisance control), and individual practices (psychic medium, writing). Several participants combine entrepreneurial work with community engagement, mentoring, or nonprofit formation.}
    \caption{Interview participants' businesses and roles. Businesses span creative industries, community organizations, service sectors, and individual practices.}
    \label{tab:participants}
\end{table}

\subsection{Follow-up interviews (N=10)}
Participants who attended a workshop were invited to opt in to a follow-up conversation.
We conducted semi-structured follow-up interviews with eight workshop participants and two entrepreneurial support personnel in the weeks following each session.
Interviews served two purposes: (1) to understand what steps participants took after the workshop based on the goals they articulated during in-session reflection activities; and (2) to offer an accountability touchpoint and additional support.
Interviews lasted 30--60 minutes, were scheduled at participants' convenience, and were conducted remotely via Zoom.
We started interviews by asking questions to situate participants' business context within the workshop (e.g., ``What are your biggest business challenges right now?'', ``What inspired you to attend the business planning workshop'') and moved towards questions to probe resilient behaviors exhibited by entrepreneurs (e.g., ``Have you asked anyone for feedback on your business plan?'').
With consent, interviews were audio-recorded and transcribed for analysis.
The protocol prompted participants to revisit their written goals, describe progress and barriers, reflect on any subsequent use of \sysname or other tools, discuss peer supports they engaged, and share ethical concerns or refusals related to AI use.
The protocol prompted participants to describe progress and barriers towards their business' goals, reflect on any subsequent use of \sysname or other tools, and consider how the workshop helped or hindered their next steps. 
We also offered resource referrals and optional check-ins where appropriate to sustain momentum.

\subsection{Informed Consent and Informed Refusal}
\begin{figure}[t!]
    \label{sec:consent-and-participants}
    \vspace{0pt}
    \centering
    \includegraphics[width=0.95\linewidth]{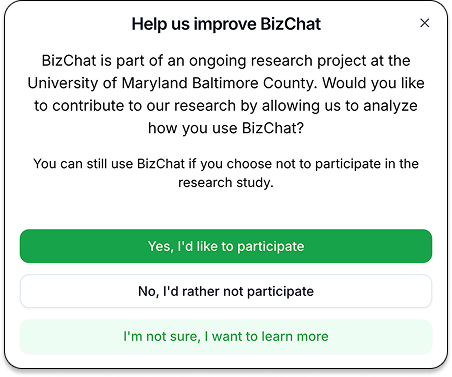}
    \Description[Consent modal screenshot]{
      Screenshot of the BizChat consent modal titled ``Help us improve BizChat.''
      The body text explains that BizChat is part of an ongoing university research project and asks whether the user would like to contribute by allowing analysis of their use.
      It also states that users can continue to use BizChat even if they do not participate in the research.
      Three clearly delineated options are presented: ``Yes, I’d like to participate'' (primary action), ``No, I’d rather not participate'' (secondary action), and ``I’m not sure, I want to learn more'' (tertiary action).
      The design operationalizes informed consent and refusal by separating tool access from research participation and offering a learn-more path for details on study purpose, data handling, and withdrawal.
    }
    \caption{We intentionally separate tool access from research participation to enable informed consent and refusal, mitigate community--academic power asymmetries, and ensure entrepreneurs can fully use \sysname regardless of research choice; the modal offers ``Opt in,'' ``Opt out,'' and ``Learn more,'' with links to plain-language details on purpose, data handling, and withdrawal.}
    \label{fig:consent-modal}
\end{figure}

We separated tool use from research participation so that entrepreneurs could access \sysname without opting into the research study.
At workshop onboarding and on the \sysname site, participants encountered a consent modal (see \Cref{fig:consent-modal}) with three clear choices: ``Opt in,'' ``Opt out,'' and ``Learn more.''
Consistent with work on informed refusal~\cite{benjamin2016informed}, ``Opt out'' allowed full use of the tool without contributing data to the study; ``Opt in'' permitted collection of de-identified interaction data and artifacts for research; and ``Learn more'' expanded details about the study, the research team, and the data collected~\cite{sum2025you}.

To mitigate power asymmetries that can fuel tensions in community–academic partnerships, we disclosed team identities and affiliations, what data (if any) would be collected under each choice, who would have access, storage location and duration, and points of contact for questions or withdrawal~\cite{sum2025you}.
Participants were reminded that non-participation would not affect access to the tool or workshop activities, and they could change their choice at any time.
If participants revoked consent, we were prepared to delete previously collected data from research storage to the extent permitted by law and platform constraints (although there was no occurrence of this).
Of the 48 users on the \sysname platform throughout the duration of the study---recruited exclusively through word-of-mouth and workshop advertising---18 chose to opt out.

\subsection{Participants}
Participants in our study included small business owners and entrepreneurs recruited through local networks and community channels. 
While recruitment was supported by the feminist makerspace, participants were not necessarily members of the space. 
The group reflected the broader entrepreneurial community connected to it.
To respect privacy and avoid re-identification within a relatively small community~\cite{abbott2019local}, we did not collect nor will report granular demographic information. 
However, participant demographics broadly reflected the inclusive orientation of the makerspace, centering women, non-binary people, and people of color in entrepreneurial practice.
More details for interview participants in particular can be found in \Cref{tab:participants}.

\newcolumntype{L}[1]{>{\raggedright\arraybackslash}p{#1}}

\subsection{Data Analysis and Evaluation}
We followed a critical incident technique to identify specific interactions or events that represented significant moments of success or challenge during the participants' interaction with \sysname~\cite{Flanagan1954CriticalIncident}. 
To this end, the first author wrote analytic memos immediately after each interview and workshop.
The notes captured conversation details, impressions, and critical incidents—such as moments of confusion, breakthrough, or successful application of suggestions—directly informing our research questions: revealing barriers and supports to adoption, illustrating how community infrastructures mediated use and refusal of AI tools, and surfacing behavioral and cognitive strategies entrepreneurs employed to build resilience.
All authors reviewed field notes, transcripts, and memos to identify and code these critical incidents. 
These codes captured different types of challenges and successes, such as moments of deep sensemaking, quick drafting of a plan, critical concerns with AI, or moments of frustration with the tool. 
Through iterative comparison and discussion among the research team, we converged on themes related to research questions (e.g., barriers to accessing capital, tension between sensemaking versus instant results, and the importance of communal scaffolds to support AI literacy and planning).

After synthesizing themes, we conducted a member check with our community partner at \prototypepgh.
In this session, we presented a concise set of emerging findings with illustrative quotes and invited critique, corrections, and elaboration.
Following guidance from community-engaged HCI practice~\cite{sum2025you}, we treated member checking as both to establish the validity of our findings and an accountability practice to further our ongoing partnership, rather than a one-off evaluation.
The refined codes form the basis of the findings presented below. 

\subsection{BizChat's In-App Data Collection}

\subsubsection{Pre and Post-Use Surveys}
For consenting BizChat users, within the interface, we administered a pre- and post-survey at two critical moments: 1) before participants generated their business plan and 2) before participants exported their business plan.
This pre-survey captured baseline measures of participants' entrepreneurial confidence and comfort, and uses cases with AI tools.
The post-survey repeated these measures and added an option for entrepreneurs to receive expert feedback from a community member on their business plan.
We report the survey and averages from responses in the \hyperref[appendix]{Appendix}.

Completion of either survey was not required to use \sysname and participants could skip it with a single button and continue directly to tool use or export. 
This design choice was intentional for opted-in participants to further mitigate power dynamics common in community-based academic partnerships, \change{although it did lead to low survey completion rates. }  
We tracked completion rates for the pre- (10 responses) and post- (4 responses) survey, and used them to complement the qualitative data analysis.

\subsubsection{Log Data}
We analyzed application logs for the workshop deployment period (January through August 2025).
We implemented our telemetry schema via the Firebase SDK, and it evolved over the deployment; when metrics required fields were introduced mid-study, we report them on the eligible subset of users for whom those fields existed and explicitly indicate this denominator (denoted by ``eligible'' in Findings).
Consistent with our consent design, opted-out users were excluded from analysis.
Primary metrics include number of users (total vs opted-in), sessions and capped session length, survey completion rates (pre among opted-in; post among pre completers), editor applies and inline-edit sessions, and AI prompt suggestion-acceptances.
For our analysis, we explicitly filtered out non-participating users, test accounts, and accounts with no business plan.
To calculate session duration, we computed capped session durations by summing consecutive event gaps within each session, capping each gap at five minutes and then averaging across sessions.

\subsection{Positionality}
\label{sec:positionality}
We situate ourselves within the community, region, and research traditions that shape this work.
We do so to acknowledge how our identities, histories, and relationships informed design choices, data collection, analysis, and claims.

The first author is an entrepreneur who grew up in the region---\change{he is a co-founder of Y-Combinator-backed technology-based, direct-to-consumer business.}
His proximity to participants geographically and familiarity with the region support rapport and trust.
He served as a facilitator during the first workshop and led the analysis.
The second author is interested in issues of power and inequity and has been conducting research in the region, engaging tangentially with the makerspace through a separate workforce development project.
\change{Similarly to the first author, he had created a tech startup in 2014.}
He approaches AI with skepticism; his involvement here stems from the last author's introduction to its potential applications in this space, though he continues to approach the technology with caution and curiosity.
The last author sought out the feminist makerspace during her PhD to find a like-minded and critical community with diverse membership and less-narrow conceptualizations of ``hacking.''
She has been connected to the community space by being first a dues-paying member, and then a formal collaborator for the last eight years.
\change{In addition, she has worked closely with over 200 entrepreneurs through her academic and industry work experience, primarily focused on the needs and aspirations of underrepresented entrepreneurs.}
Outside of research, her creative pursuits include professional dance, which further connects her to the makerspace's diverse forms of making and expression.

\change{Our backgrounds enabled rapport and helped position us as relative insiders who understand the pressures of business planning and capital access firsthand. However, our academic positions mean we do not share the financial precarity that shapes necessity-driven entrepreneurship for many participants.}
\section{Findings}
\label{sec:findings}
In this section, we detail how \sysname helped entrepreneurs translate ideas into the ``language of business,'' quickly generate first drafts, and edit plans in their own voice. Yet, this speed sometimes bypassed deeper business sensemaking, surfacing a tension between efficiency and reflection. In addition, we share how participants turned to peer networks to refine their plans, build AI literacies, and seed their own community infrastructures to negotiate adoption.

\subsection{Leveling Barriers to Accessing Capital and the ``Language of Business''}
\label{sec:language-of-business}
Overall, participants viewed \sysname as a tool that levels the playing field for accessing capital by: (1) translating entrepreneurs' ideas into the ``language of business'' and (2) shortening the time it takes to create a first draft.

Participants valued that \sysname,  ``translated everything to the right speak'' (P1).
P2 and P3, who are starting a makerspace for vocational training for neuro-divergent individuals, shared how \sysname reflected their words back in recognizable terms: ``and then to have it regurgitated back to me in like business lingo, which is not my expertise was like, thank you. ...that's why I really enjoyed it.''  
P2 framed this translation as navigating a gate-keeping language that previously felt inaccessible:
\textit{``We are not business people, but that levels the field. So if someone has an idea like, I've got this great idea, but I'm not a business person, if you have the AI generators ...it's a hurdle that somebody doesn't have to take, you know?''}
For P2, \textit{``hurdles''} meant taking unstructured thoughts from \textit{``writing whiteboards''} and \textit{``typing pages''} into polished formats.
This was reflected in usage logs: across 13 eligible users' onboarding conversations, participants engaged in an average of 6.54 user turns (median 7, max 12), and crucially, they clicked \textit{``Not sure? Click for some ideas''} (see \cref{fig:bizchat-onboarding}.1) prompt-suggestions 3.27 times per active user---nearly half the median onboarding chat-turns. 
This shows that during onboarding, participants leaned heavily on AI-generated suggestions to surface ideas and translate them into business language, a concrete instantiation of \sysname's low-floor, high-ceiling design that helped them articulate what they meant even when they were unsure.

For P4, \sysname lowered the floor by quickly consolidating information for her.
P4 shared how gathering previously scattered business information from her website (e.g., product line, mission statement copy) reduced the effort needed to assemble her plan and lessened the stress of tracking down and recalling those details on her own. 
P4, who self-describes as a \textit{``grant-queen''}, but \textit{``not a techie''} now felt empowered to take the information from \sysname, and use it for other grant applications: \textit{``The information that I learned from the \sysname, I now can take pieces of that... and then just pop it in, instead of looking through all my papers.''}

Participants also valued the agency \sysname, unlike generic text interfaces with un-editable outputs, afforded them to reshape that \textit{``right speak.''}
As P4 put it, \textit{``It allowed me to see it, and in sections, and then I can work with it that way ...I needed to reword it in a way that sounds like I wrote it, and not a robot wrote it''} (P4).
But, building the agency to edit the business plan was not purely due to interface design, but also in how the research team and community partner positioned business plans as living, evolving documents during the workshop series.
For P2, this was a major un-blocker, \textit{``I didn’t know the plan could be flexible ...I thought once you had a plan, it was carved in stone.''}
This framing in the workshop was a helpful affirmation that P2 could adapt her plan as her business adapted.

In addition to translating entrepreneurs' ideas into the language of business, participants also expressed how \sysname significantly reduced their time to get a first draft.
Both P4 and P5 reported the time it took to draft a business plan went from \textit{``six months''} to a matter of \textit{``an hour''} (P4) or \textit{``minutes''} (P5).
As P5 put it, \textit{``It took me six months to write my last business plan... ...[to make sure] all the I’s were dotted, all the T’s were crossed.''}
For P5, who runs a custom jewelry and apparel business, building this momentum to get a first draft had a very tangible outcome. 
Later, she reported using the business plan \sysname helped her draft to apply, and ultimately get accepted to a major international fashion event. 
Even if the draft was imperfect, \sysname provided the momentum she needed---something tangible to share with a friend---which ultimately set her on the path toward achieving a major milestone for her business.

For P6, who self-identified with ADHD, speeding up the drafting process intersected with equity.
Unlike domain-agnostic tools, \sysname provided a non-blank starting point: \textit{``I have ADHD... it was super helpful... it’s hard to get from zero to something going''} (P6).
She described the tool as moving her \textit{``to a starting point that’s like further down'' and as ``add[ing] like the equity element''} toward parity with those without ADHD (P6).



\change{For P7, reducing the time to create a first draft helped reframe the emotional stakes of sharing his business plan.
Previously, P7 viewed sharing his business plan as exposing himself to the risk of IP theft from possible lenders. 
For instance, P7 recalled this anxiety around sharing plans: \textit{``If you turn me down, you have my plan... you can just give it to your kid, and they can go off and be rich... I just handed it to you on a piece of paper. Here you go! Here's all my hard work''}.
P7, reflecting on his past experiences as a member of a marginalized community, described how sharing his business plan once carried a form of extraction anxiety---the fear that lenders could reject him yet still profit from his hard work in creating the plan.
When business planning required months of painstaking effort, sharing that plan felt like handing over deeply personal labor that could be exploited.
By reducing the time and emotional investment required to produce a draft, BizChat shifted what felt vulnerable: the business idea itself remained valuable, but the document representing it no longer carried the weight of months of personal struggle.
Importantly, this shift reflects a change in participants' \textit{subjective assessment of risk}, rather than any change in the objective legal ownership or factual risk of IP theft.
As P7 explained, when a draft takes an hour rather than six months to create, the anxiety of potential theft diminishes---not because the IP is less valuable, but because the personal cost of producing the artifact to share is dramatically lower.
Moreover, the ease of iteration meant P7 could quickly generate updated versions or tailor plans for different audiences, reducing the sense that any single document represented an irreplaceable, static record of his ideas that could be stolen and exploited.}

\sysname helped entrepreneurs overcome barriers by translating their ideas into the \textit{``language of business''} and accelerated their drafting process, while still supporting agency in shaping their plans.
Yet, while participants valued these efficiencies, they also grappled with the balance of quickly drafting a plan, and engaging more deeply in making sense of their business.
Thus, in the next section we highlight this tension.

\subsection{Business Sensemaking vs. Instant Results}
Entrepreneurs (P4, P5) were blunt that time is scarce and business plans are boring.
The pull to get an instant draft and move on was real and reasonable, especially given time and resource constraints.
However, in the context of our resilience planning workshops and deployment of \sysname, this surfaced a tension between the need for instant results, while supporting long-term planning and business sensemaking.

As mentioned in the previous section, quick results were celebrated because they removed the drudgery of business planning.
As P4 put it, \textit{``I don’t have time to sit down and write out a business plan... it’s so freaking boring to create step by step.''}
For P4, who had previously spent six months creating a business plan, the instant results of the \sysname workshop, provided the necessary momentum to get over the boredom of business plan writing.

However, entrepreneurial support personnel were keen to share how quick drafting can mask gaps in preparedness.
As P8 put it, \textit{``If [\sysname] makes a plan that helps people get money, [but] they have no business owning a business and they don’t have the skills, [the plan] doesn't help.''} (P8)
She also cautioned against over-trusting tidy output: \textit{``One of the issues, where [\sysname] can’t find the information on the website: Is it is making something up? It sounds compelling, so people are not going to challenge it.''}

Through low-floor high-ceiling design principles, \sysname started, but failed at fully navigating this tension.
Through onboarding, entrepreneurs are asked to think through many aspects of their business that ultimately appear in their business plan. 
P8 highlighted how this can be powerful for self-reflection, \textit{``If they survive the gauntlet of being asked all of the questions, that could be really helpful to people.''}
P8 expanded to say that, in being asked these questions, the entrepreneur could have realizations about themselves---what if they realize they actually do not want to start a business?

Participants also described onboarding as just-in-time learning that turns unknown unknowns into known questions.
Contrasting with general-purpose AI tools like ChatGPT, P6 noted the importance of \sysname's guided flow over open-ended prompting: \textit{``with ChatGPT, it feels like I’m guiding the tool, whereas like \sysname feels like it’s guiding me.''}
P6 added: \textit{``ChatGPT ...it takes more brain power to even know what to ask it.''}
Building on that, P2 shared that \textit{``It's telling me what I don't know, when I don't know.''}
P3 emphasized how the tool surfaced blind spots, describing it as something that \textit{``takes everything I’ve learned and polishes it and clarifies it and make sure I think about things that maybe I hadn’t thought about.''} She linked this reflection to a broader goal of staying competitive, noting that \textit{``the tools can help you learn about what the questions are, so that you stay competitive and that you stay current.''}



\change{Beyond surfacing unknown unknowns, for P6, the scaffolding was a helpful tool to shift away from ``shyness'' and get her in the mindset of running her business: ``I'm someone who shies away from dreaming bigger...It helped me get in the mindset of running my business.''
For P6, the structured prompts asking her to articulate her target market, value proposition, and growth strategies forced her to inhabit the role of a business owner—thinking strategically rather than operationally.
By centering her expertise and asking her to make decisions about her business throughout the onboarding process, BizChat positioned her as the authority, which shifted her self-perception from someone "dreaming" to someone actively planning and executing.
P9, who once went through the \prototypepgh Incubator program and now runs communications at a local makerspace, observed this pattern across many entrepreneurs, questioning whether the biggest barrier for entrepreneurs is less about technical knowledge and more about developing the entrepreneurial mindset: ``Do people need more training in having an entrepreneurial spirit?''
P9's question highlights a critical tension: while BizChat's scaffolding helped P6 adopt a strategic mindset by prompting her to make concrete business decisions, it remained unclear whether this was genuine mindset development or merely performing the role of entrepreneur through AI-prompted responses.
P9's skepticism—echoed in her earlier concern about whether AI-generated plans truly "train" entrepreneurs—suggests that quickly producing a polished plan might substitute for, rather than cultivate, the deeper entrepreneurial thinking that comes from wrestling with business challenges themselves.
In this way, \sysname's scaffold encouraged reflection, learning, and goal setting, yet the question of whether AI-assisted planning builds or bypasses entrepreneurial capacity remained unresolved.}

For entrepreneurs without a website, \sysname asks questions about each section of their business plan before creating a draft.
But if an entrepreneur’s website already contains enough relevant information, \sysname skips these questions and generates a draft immediately.
In this case, the entrepreneur may not encounter the \textit{``gauntlet of questions''} that P8 describes.
P1’s case illustrates this miss.
He arrived with a website, so \sysname produced a complete plan immediately.
What drew him to the workshop, however, was the promise of resilience planning as an ongoing practice, not just a document.
He warned that over-ease can reduce ownership: ``if you make things too easy, people don’t value it enough... if you get to the finish line, you have less ownership over it.''
P1 had recently transitioned back to running a freelance design agency after a round of layoffs at his previous employer. For an operator with fifteen years in business who wanted to think through the right contingencies, \sysname's low-floor and high-ceiling approach did not sufficiently slow down or deepen reflection.
Log data tells a similar story.
After onboarding, users did not significantly iterate on their business plan, with users (20 eligible) averaging 0.77 rich-text edits (see \Cref{fig:bizchat-overview}d) and 0.77 AI-applies per user (see \Cref{fig:bizchat-overview}b). 
Yet, despite limited post-onboarding edits, users (26 eligible) averaged a session length of 32 minutes, with a median of seven onboarding chat-turns---suggesting users spent substantial time reflecting and engaging during the guided onboarding flow, even if they did not edit their business plan after.


While instant results provided necessary momentum and lessened the drudgery of planning, they also risked masking blind spots. 
The tool helped to uncover unknown unknowns and foster reflection, yet in cases like P1's reveal risks when drafting outpaces deeper engagement. 
Both within and beyond our workshop series, entrepreneurs turned to peer support to develop AI literacies to navigate this tension, which we detail in the next section.

\subsection{Communal Scaffolds for AI Literacy and Resilience Planning}
\label{sec:communal-scaffolds}
After the workshop, peer support helped to take participants' business plans to the finish line.
For P5, continued peer support after the \sysname workshop was crucial to turning her plan into a final draft.
She recalled how a friend helped stitch together financial projections to finalize her plan for an application, and she ultimately used the \sysname-started plan to apply and get accepted to a major international fashion event.
Other workshop participants, P6 and another, met up bi-weekly to continue refining their plans together.

While some participants described ad-hoc meetups to take next steps, others wanted to turn this into more formal community infrastructure.
For instance, both P7 and P4 suggested hosting their own variations of the \sysname workshop within their local communities.
\change{P4 expressed interest in hosting her own version of the \sysname workshop within her immediate social circle, describing plans to run an informal, small-group session out of her garage to share the tool and process with friends who were curious but hesitant to engage through formal programs.
For P4, who emphasized the importance of in-person contact and relational support during the workshop, it mattered that {\em she} would be the one creating this intimate environment.
She envisioned hosting in a way that would foster connection and comfort: \textit{``When I do it, I want to be able to feed the people while we're there. But I want to keep it like to 10 to 15 people, because I think that's a good size.''} (P4).}

P7, who mentors young adults through a federal workforce training program and runs a nuisance control business, explicitly wanted to host \sysname workshops in places like his local public library and local entrepreneurial hub with small cohorts. 
He described this desire as part of a goal towards supporting greater AI-literacy in his community, likening it to recipe cards: \textit{`I want to introduce literacy with AI—like Betty Crocker sent out recipe cards.''}
But, here, for P7, {\em how} AI is introduced mattered.
He contrasted \sysname's workshops with other academic workshops, in the sense that it disseminated digital literacy at a point focused on practical application rather than new technology:
\textit{``What I liked… was `How can I help someone who obviously needs help right away'... it wasn’t like some `We’re gonna help you start a business' type of thing.''}
For P7, as he plans to bring this to his own community, they emphasized the need to have a trusted messenger.
He wanted to host the workshop himself because \textit{``If I did it, they’d believe it.''} 
Here, P7 signals more than personal pride---he emphasized that, in his view, AI literacy is relational.
P7 framed information sharing as an act of trust, where credibility depends on {\em who} cross the boundary to bring back new knowledge.
P7 described this in the metaphor of ``going to get the bread,'' drawing on histories of someone risking themselves to bring resources from inside slave-owners homes back to the field.
In the context of AI, they see themselves as the one who must ``run in, grab the information, and run back out,'' testing new tools before introducing them to others.

P7 grounded this in Pittsburgh's geography and culture, reflecting on how a local main street acts as both a physical and symbolic divider. \textit{``Some people will say it’s the train tracks. Which, in fairness, it is.''} they noted, pointing out how these divides shape whose knowledge is trusted. 
P7 shared how his tattoos of a Pittsburgh landmark were his proof of credibility—his way of signaling, \textit{``I’m from here, I’m one of you.''} 
Here, his comment that \textit{``if I did it they’d believe it''} underlines that information from \textit{``across the tracks''} may not carry the same weight, as if they were the messenger.

Despite P7 wanting to co-opt the \sysname workshop for his own community, our intentional framing of \sysname as a business planning workshop, rather than an AI workshop, created tension.
For instance, P9 saw us not directly calling out AI in advertising materials, as normalizing use of the technology:
\textit{``I do think there's a value in being awake to the ways things are kind of slowly normalized ... when complete abstaining doesn't happen, does it set the stage for normalizing the use of AI in other places?''}
P9, who self-describes as \textit{``not a fan of AI''}, and has \textit{``never even opened ChatGPT,''} for \textit{``reasons, political, environmental, [and] creative''}, views this normalization as harmful.
Thus, in P9's view, not clearly centering the use of AI in the tool, removed agency from entrepreneurs to engage in refusal of use.

While participants like P4 and P7 articulated intentions to act as trusted messengers introducing \sysname within their communities, this study does not observe whether such introductions lead to subsequent community adoption or sustained AI use.
Taken together, these accounts show how \sysname workshops seeded peer-led infrastructures around business development and AI adoption within the entrepreneurs we interviewed.
Yet, as participants considered hosting or adapting the workshops, questions emerged about {\em how} AI should be introduced, {\em who} should introduce it, and whether this normalization of use itself is problematic.

\section{Discussion}
\change{Our findings reveal that resilience emerged from the interaction between \sysname, community infrastructure, and peer relationships.
We observed this in three interconnected ways. 
First, entrepreneurs transformed their uncertainty about AI---what P3 noted as ``things that maybe I hadn't thought about''---into addressable questions. 
What began as fear of falling behind became specific, actionable gaps in their business plans. 
Second, participants built peer-led infrastructures that extended beyond our intervention. P5 needed a friend to complete financial projections. 
P6 met bi-weekly with another participant to refine plans. 
P7 and P4 independently planned to host workshops introducing BizChat in their own communities.
Third, participants collectively negotiated when to adopt, adapt, or refuse AI outputs, where they developed shared strategies for editing, contextualizing, and evaluating them. Critically, this included legitimizing refusal as informed choice as debates about AI's appropriate role (exemplified by P9's transparency concerns) became sites for collective sensemaking.
These observations ground the design implications that we discuss below. 
}

\subsection{Designing for Resilience}
\change{The forms of resilience above---transforming uncertainty into actionable questions, building peer infrastructure, collectively negotiating AI use---required conditions that our deployment partially created.
Reflecting on our experience, we surface three design implications for intentionally supporting such resilience: productive friction to scaffold reflection, communal scaffolds to support peer-based AI literacy development, and co-optability to enable communities to appropriate tools on their own terms.
}

\subsubsection{Productive Friction}
Previously, we shared how while \sysname aided participants by leveling barriers to accessing capital and the ``language of business,'' participants often traded their own reflection with instant drafting.
In this way, \sysname's low-floor, high-ceiling design principle was successful in lowering barriers to creating business plans, moving from ``months to minutes.''
However, our log data showed that most entrepreneurs did not iterate significantly on their business plan.
This suggests that many participants are settling for ``good enough'' AI outputs without deeper engagement. 
This behavior stands adjacent to prior work in HCI, which frames AI-resilient interfaces as those that support users in noticing, judging, and recovering from AI errors~\cite{glassman2024ai}. 
The passive acceptance and the absence of friction or pause for users to question the AI-generated content may risk undermining resilience-building practices, as such reflections are central in creating a strong adaptive capacity \cite{BUCKNELL2022111234}.

We argue for enabling ``productive friction'' --- moments of intentional pause that prompt users to reflect, revise, or make sense of AI outputs~\cite{kapur2008productive, glassman2024ai}.
We draw inspiration from learning sciences literature that advocates for productive struggle and failure in a supportive environment to foster deeper engagement and skill-building~\cite{kapur2008productive, murdoch2020feeling, warshauer2015productive}.
This idea also aligns with the notion of seamful design \cite{inman2019beautiful} and design friction \cite{haliburton2024longitudinal} within HCI, both of which posit the benefit of sometimes creating friction in the experience to deepen reflection and understanding.
Taking these ideas together, we advocate for productive friction, which entails building into the design deliberate difficulty that helps users to understand and improve their judgment of the system's output.

\sysname partially operationalized productive friction through the onboarding flow, where participants were asked a series of detailed questions about their business before receiving AI-generated outputs.
This served as a moment of reflection before the AI gave out the results. 
Reflecting on our experience, we note that deciding where and how to introduce productive friction is a power-laden design choice.
It encodes assumptions about what kinds of struggles are valuable. 
Rather than predefining where users must slow down, designers should offer ways for users to signal whether they seek quick drafting or deeper reflection, and align friction with their goals.
Building from work in CBPR, which emphasizes participant agency and cautions against paternalistic design choices (e.g., \cite{gautam2020p, vines2013configuring, arnstein1969ladder}), we argue that this friction should be \textit{negotiable} as a key consideration is who \emph{wants} to engage in the productive friction and who \emph{needs} quick results. 
For entrepreneurs facing a large number of tasks and multiple demands on their time, mandatory friction can be exclusionary. 
We advocate for designing such that users have a choice when to deeply engage and when to prioritize speed. 

In \sysname, we implemented this negotiability in the onboarding flow (\Cref{fig:bizchat-onboarding}) by allowing users to skip it completely. 
Yet, in practice, most chose \emph{not} to skip it and went through the detailed onboarding. Some even welcomed the friction. 
For example, P1, who initially skipped the onboarding flow, later expressed disappointment because that reflection via business planning was what he was seeking.
Tied to our low-floor, high-ceiling design principle, we argue that enabling negotiable friction will allow users who value reflection and if their context allows, will engage with the friction, whereas those who need quick output can proceed without being blocked.

While designing for \textit{negotiable productive friction} helps respect user agency, it has the risk of users consistently choosing speed over reflection, and thereby building over-reliance on AI. 
Building on \citet{glassman2024ai}, we argue that systems (or the context of their deployment) must scaffold noticing, judging, and recovering practices with the output to promote AI literacy: helping users recognize when an output is ``good enough,'' when it should be challenged, and how to act on either decision. 
\change{Future work could explore how to design friction points that help users notice and recover from LLM-specific errors, such as hallucinated details or generic outputs that do not reflect their actual business context.}
As we highlighted in \Cref{sec:communal-scaffolds}, in practice, \sysname users turned to peers to support iterating on AI-generated drafts, and taking next steps with their business plan.
Building on this observation, we turn to our next design implication.

\subsubsection{Communal Scaffolds}
As we touched on above, low-floor, high-ceiling design made it possible for entrepreneurs with diverse digital skills to draft business plans, but it was the communal context that ultimately enabled them to exercise agency, negotiate with AI outputs, and build AI literacy together.
In our deployment, lowering the barrier to generating a business plan created opportunities for help-seeking and peer collaboration. 
For example, after using \sysname, participants met with peers over coffee or refined their plans with mentors. 
It gave them a concrete reason to reach out.
\sysname also became a vehicle for supporting AI literacy: two participants independently sought to introduce the tool to their own communities, with P4 explicitly framing this as a way to teach others how to use AI for business planning.
Reflecting on these experiences, we see the value of carefully deployed AI systems to enable users in developing AI literacy, collective sensemaking, and negotiating the meaning of system outputs. 

We argue that AI systems should be designed and deployed to amplify the communal scaffolds already present in the context. 
\change{In our case, for example, P5 could generate a draft with BizChat but needed friend's domain expertise to build the financial projections her grant application. 
Looking at the output, P4 valued \sysname's translation into ``business language'' but sought peer feedback to make it sound like her, which was an authenticity judgement that benefits from others who know her voice.}
This is a move to support users to construct an understanding of AI and its outputs together; a way to overcome individual limitations and build collective resilience. 
Prior research in HCI and entrepreneurship context supports the importance of building such communal scaffolding and capacity building~\cite{dillahunt2014fostering, kotturi2022tech, kotturi2024sustaining, romero2024exploring, dillahunt2025development}.
\change{More importantly, we note that individual judgment about AI outputs is fragile, especially as systems evolve rapidly. But when communities develop shared heuristics for engaging with the system or evaluating its outputs, they build collective adaptive capacity that persists beyond any single tool. AI literacy becomes a community resource, transforming individual vulnerability into collective resilience~\cite{dillahunt2025development}. This aligns with prior HCI research.}
For instance, Kotturi et al. demonstrate that introducing AI tools in community contexts can reduce fear of being left behind and create opportunities for collective learning~\cite{kotturi2024deconstructing}, while other work shows  entrepreneurs socially negotiate norms around AI adoption, underscoring that decisions about what counts as appropriate use are rarely made in isolation~\cite{romero2024exploring}.
We extend this by arguing that systems should actively support these communal processes. 
In the context of AI-supported business planning, this means treating AI literacy as socially constructed and context-specific, rather than as a fixed curriculum to be delivered to individuals. 

\change{
Treating AI literacy as socially constructed also means recognizing that communities must be able to collectively choose not to adopt, that is, to refuse.
Feminist scholarship on technology refusal, such as Manifest-No (a collective statement against extractive AI systems~\cite{manifestno}) and SUPERRR Lab's statement on AI (articulating concerns from intersectional feminist and ecological justice perspectives~\cite{superrr2024}), positions collective refusal as an act of agency and political resistance to harmful technologies, extending HCI's view of refusal as an informed stance when technologies misalign with one's goals, values, or constraints~\cite{satchell2009beyond, selwyn2006digital, baumer2014refusing, gautam2024reconfiguring}.
In our study, refusal operated at two levels: formal refusal to participate in data collection (only 30 of 48 users opted in) and practical refusal of AI use for specific tasks.
P9 exemplified the latter, describing herself as "not a fan of AI" for "political, environmental, [and] creative" reasons and having "never even opened ChatGPT."
However, our methodological choice to de-center AI in workshop framing may have inadvertently made informed refusal harder---by not explicitly naming AI's role upfront, we risked obscuring what participants were choosing to engage with or refuse.
This tension highlights that designing for resilience includes designing for informed refusal by making the stakes of AI adoption visible enough for communities to critically evaluate when and how these tools fit their needs~\cite{kotturi2024deconstructing}.
}

\change{\sysname operationalized these principles by positioning business plans as drafts, encouraging participants to share, revise, and reflect with others—turning AI use into a site of communal sensemaking where adoption, adaptation, and refusal could all be negotiated collectively.
Thus, we call for designing AI systems that intentionally scaffold community-based resilience practices: by positioning outputs as negotiable, making collaboration and feedback around AI-generated content easy, and creating affordances for groups (not just individuals) to build shared capacity and literacy together.}

However, enabling these communal scaffolds is not enough on its own. 
They must also be supported by systems that communities can appropriate and use on their own terms. 
Without control over how the system fits into their practices, communities may struggle to sustain engagement or build confidence in the tool. 
Thus, we turn next to the challenge of designing for co-optability, exploring how systems can support forms of ownership beyond technical maintenance and enable communities to take ownership of the processes that shape their use.

\subsubsection{Co-optability}
\change{In our deployment, P4 and P7 expressed desire to bring \sysname into their own communities; P7 to his library and entrepreneurial hub, P4 to fellow entrepreneurs she mentors. 
They valued the system. 
However, neither wanted to simply reuse the tool. 
They wanted control over how it was introduced, who was in the room, and how it was framed. 
At the same time, they were aware of the gap in their technical skills and infrastructural resources to run and maintain it.}
This reflects a form of ownership over the process, one that goes beyond direct platform stewardship. 
We ask, how can we support the community to have such ownership? 

We argue for emphasizing \emph{ownership of processes over ownership of the product}.
Communities should not be expected to host servers or manage GitHub issues if these are not relevant goals for them, but they can exercise ownership in how the tools are introduced, adapted, and integrated locally.
P7’s comment that \textit{``if I did it, they’d believe it''} illustrates one way process ownership can be tied to messenger credibility and trust; who delivers the technology matters. 
He likened introducing new tools to \textit{``going to get the bread,''} which, in his framing, positioned knowledge-sharing as carrying risk and responsibility, and suggested that a trusted messenger may be key to ensuring the message is received.
If systems cannot be reframed and delivered by these local messengers, they may risk being perceived as \textit{``from across the tracks''} and, as a result, struggle to take root.
\change{This matters for resilience because tools introduced by outsiders risk being abandoned when external support ends. When communities own the process of introduction and adaptation, they build capacity to sustain and evolve their use over time, even as specific tools change~\cite{kotturi2024sustaining}.}

Designing for co-optability, then, means creating systems that make it easy for communities to take and make the tool their own.
In practice for \sysname and other generative AI tools, this could include lightweight administrator roles that allow local facilitators to configure onboarding questions, adjust framing to match local language, and provide just-in-time support for \textit{``laundry list''} operational skills (e.g., password management, account setup) that often stand as hidden barriers to adoption.
Beyond technical access, systems can offer ways for facilitators to submit feature requests, bug reports, and workshop feedback that directly integrate into maintainers’ workflows, enabling a model where the community helps shape how the system is maintained and how maintenance is prioritized.
\change{Future work could explore interfaces that allow community facilitators to configure model selection and communicate tradeoffs, such as privacy, cost, and accuracy, to participants.}

\subsection{Methodological Implications: Deployment Details and Reflections}
We chose business planning as a context for resilience building because it is a practice that facilitates reflection, can yield tangible results (e.g., access to capital), and where entrepreneurs can have expertise over the outputs (see \Cref{sec:context-for-resilience-building}).
We explicitly framed \sysname's outputs as drafts rather than finished products to encourage help-seeking---in service of our broader framing of supporting resilience.
Beyond the tool's design, our deployment methodology was also aligned with a resilience orientation.
Given that accountability is an ongoing process, and following critical work in HCI which emphasizes participant-researcher reciprocity, we explicitly positioned follow-up interviews as accountability check-ins for entrepreneurs, in service of progress on their business plan.

Another way we foregrounded resilience was by explicitly advertising and framing the workshops as ``Resilience and Uncertainty Planning Workshops,'' making participants' business goals the focal point of the sessions, rather than the presence of AI.
We adopted this framing as a way to create an accessible entry point into the workshops---participants were not expected to already be comfortable with AI---and to avoid attracting those only motivated by the novelty of generative AI.
This decision succeeded in reaching entrepreneurs focused on business planning.

However, it also introduced tensions around transparency.
Because we de-emphasized AI in our outreach materials, several participants shared that it was not immediately clear that AI would be part of the workshop.
One participant (P9) explicitly raised concerns about this lack of transparency, highlighting a dangerous tension: our framing, while inviting broader participation, risked eroding trust with community collaborators. \change{Future deployments should make AI's role explicit upfront, even when de-centering it pedagogically.}

Given that many commercial AI platforms retain user data to train models, our opaqueness around the use of AI also conflicted with IP concerns among entrepreneurs.
These concerns created an important methodological and operational constraint for our deployment, motivating us to reflect carefully on which models we used and why.
Throughout the two-year development of \sysname, there were no standardized APIs for LLMs. 
While we initially implemented on top of open-source models and a private server, their rapid improvement cycles made it difficult for our research team to keep pace, and hosting them at comparable latency to commercial APIs was costly. 
Given the small scale and local nature of our deployment, it was not operationally feasible to self-host models.
Today, with the emergence of standardized API specifications (e.g., the OpenAI SDK spec~\cite{openai-openapi}) and routing tools like OpenRouter~\cite{OpenRouter} that allow users to select models through a common interface, there is a more feasible path to balancing this tension: supporting rapid iteration on a small team while reducing dependence on state-of-the-art commercial providers.
However, each platform still maintains its own data retention and model training policies, creating uncertainty for participants and also for us. 
For small research teams, even hosting their own infrastructure for open-source models can be prohibitively expensive---both in compute costs and in the expertise required to maintain production-level reliability. 
We argue that this highlights a broader infrastructural gap in HCI research: there are few shared, privacy-preserving resources that enable researchers to deploy cutting-edge models without defaulting to commercial APIs. 
Creating such a resource pool would lower the barrier for small teams, making it possible to both serve state-of-the-art models and protect participants’ data. 
As more researchers investigate small-scale deployments, where the cost and infrastructural trade-offs of serving open-source models are prohibitive, addressing this gap could make experimentation more equitable, reproducible, and accessible.
In practice, this could be implemented on top of existing research infrastructures like NSF CloudBank~\cite{CloudBank}.
Building from \textit{productive friction} and \textit{communal scaffolds for AI literacy}, in future deployments of \sysname, as we introduce more open source models, productive friction around model selection could be another point where communities negotiate tradeoffs and support AI literacy.

\subsection{Limitations}
There were several limitations to this study. 
First, we did not conduct formal evaluations of business plan quality, partly due to leadership turnover among \prototype staff. 
While we mitigated this through ongoing accountability check-ins, this limited our ability to assess long-term impact and formally evaluate the generated business plans. 
\change{
Future work can develop evaluation approaches that assess whether AI systems support entrepreneurial agency rather than merely producing professionally formatted documents.
For instance, one promising direction is to evaluate whether key business details that entrepreneurs articulate during onboarding chat conversations are preserved rather than summarized out---a frequent occurrence~\cite{goyal2022news}---in the final generated plan.
Such an approach would assess not just output quality, but whether the system successfully supports entrepreneurs in maintaining agency and specificity over their business narrative. 
Ultimately, as business plans serve as both a reflective scaffold and document to access capital, longer-term evaluations could measure how well business plans supported these two ends as they are relevant to a particular BizChat user's goals.
}

Second, as with many community-centered deployments, our findings are shaped by the specific context and relationships in which \sysname was embedded. 
The socio-economic dynamics, including community-level trust and support structures for entrepreneurs, can differ in other contexts. 
Additionally, while participants such as P4 and P7 expressed intentions to act as trusted messengers introducing \sysname within their communities, this study did not observe whether such introductions led to subsequent community adoption or sustained AI use.
Future work should empirically examine how messenger credibility, local trust, and community identity shape whether AI tools are taken up, adapted, or refused over time.

Third, our interview sample size was small (10 entrepreneurs and support personnel), and participation was opt-in, introducing self-selection bias that likely reflects who felt empowered or resourced to engage with an AI tool. 
\change{Finally, survey completion rates were low, limiting our ability to draw quantitative conclusions about changes in entrepreneurial confidence or AI comfort—though qualitative interviews provided richer insights into these dynamics.
In future work, we will investigate balancing the need for survey completion with reducing extractive data collection practices, such as by communicating more clearly to users the purpose of surveys for improvement of the system and support of entrepreneurs more broadly.}
Thus, the findings reported here may not be generalizable; however, the lessons learned from the design and deployment can be transferable to other contexts and inform researchers interested in working with entrepreneurs or in fostering resilience amid rapid technlogical change. 

\section{Conclusion}
In this paper, we introduced \sysname to a feminist makerspace and entrepreneurial hub in Pittsburgh, Pennsylvania.
In doing so, we moved beyond asking whether generative AI belongs in entrepreneurial practice to exploring what happens when it is woven into existing business and communal workflows.
We showed how a community-centered design and deployment approach not only lowered technical barriers but also strengthened the social scaffolds that enable entrepreneurs to build resilience.
\change{Our findings revealed a central tension: while \sysname reduced barriers to accessing capital by translating ideas into "business language," this ease raised questions about whether instant AI outputs undermine the reflective sensemaking essential to planning.
Entrepreneurs navigated this tension collectively—transforming uncertainty about AI into shared strategies for evaluating, contextualizing, and when appropriate, refusing outputs.}
AI is not the first technology to exacerbate inequalities in adoption, and it certainly will not be the last.
In designing for resilience, we argue that the real challenge is designing through the hype toward practical, situated use.
Amid the hype of each new innovation, we are tempted to ask how our practices around design should change.
But as our findings show, it is human resilience and community infrastructure that persist, and our task as designers is to respect and reinforce those systems.

\begin{acks}
First and foremost, we thank our participants and the Prototype PGH community. 
This work was funded, in part, by UMBC's Alex. Brown Center for Entrepreneurship.
Generative AI tools assisted in the preparation of this work: 
ChatGPT (GPT 5.1) and Claude (Sonnet 4.5) were used for summarization and refinement of text; Gemini's Nano Banana was used to generate assets for Figure~\ref{fig:teaser}. 
\end{acks}
\appendix
\label{appendix}

\section{Survey Questions and Results}
Participants responded to a pre-survey ($n{=}10$) and post-survey ($n{=}4$) measuring clarity, uncertainty, and AI comfort. Average Likert-scale scores (1--5) are reported below with change ($\Delta$) computed as Post $-$ Pre.

\begin{itemize}
    \item \textbf{Clarity on Steps:} How clear are you on what steps your business needs to take to move forward?  
    \textit{Pre: 3.32, Post: 3.25 ($\Delta = -0.07$)}
    
    \item \textbf{Uncertainty (Strategic):} I am uncertain about the \emph{strategic direction} of my business.  
    \textit{Pre: 3.34, Post: 3.00 ($\Delta = -0.34$)}
    
    \item \textbf{Uncertainty (Technical):} I am uncertain about the \emph{technical aspects} of my business.  
    \textit{Pre: 3.18, Post: 3.00 ($\Delta = -0.18$)}
    
    \item \textbf{Uncertainty (Marketing):} I am uncertain about the \emph{marketing aspects} of my business.  
    \textit{Pre: 2.87, Post: 3.00 ($\Delta = +0.13$)}
    
    \item \textbf{Uncertainty (Operations):} I am uncertain about managing the \emph{operations} of my business.  
    \textit{Pre: 3.96, Post: 3.00 ($\Delta = -0.96$)}
    
    \item \textbf{Uncertainty (Support):} I am uncertain where to seek trustworthy \emph{social support and guidance}.  
    \textit{Pre: 3.81, Post: 3.00 ($\Delta = -0.81$)}
    
    \item \textbf{AI Comfort:} I feel comfortable using AI tools in my business planning.  
    \textit{Pre: 3.54, Post: 3.00 ($\Delta = -0.54$)}
\end{itemize}

\bibliographystyle{ACM-Reference-Format}
\bibliography{main}

@article{kotturi2024sustaining,
  title={Sustaining Community-Based Research in Computing: Lessons from Two Tech Capacity Building Initiatives for Local Businesses},
  author={Kotturi, Yasmine and Hui, Julie and Johnson, TJ and Sanifu, Lutalo and Dillahunt, Tawanna R},
  journal={Proceedings of the ACM on Human-Computer Interaction},
  volume={8},
  number={CSCW1},
  pages={1--31},
  year={2024},
  publisher={ACM New York, NY, USA}
}

@article{termeer2013organizational,
  title={Organizational conditions for dealing with the unknown unknown: Illustrated by how a Dutch water management authority is preparing for climate change},
  author={Termeer, Catrien JAM and van den Brink, Margo A},
  journal={Public Management Review},
  volume={15},
  number={1},
  pages={43--62},
  year={2013},
  publisher={Taylor \& Francis}
}

@book{weick1995sensemaking,
  title={Sensemaking in organizations},
  author={Weick, Karl E and Weick, Karl E},
  volume={3},
  number={10.1002},
  year={1995},
  publisher={Sage publications Thousand Oaks, CA}
}

@inproceedings{parker2025participatory,
  title={Participatory AI Considerations for Advancing Racial Health Equity},
  author={Parker, Andrea G and Vardoulakis, Laura M and Alla, Jatin and Harrington, Christina N},
  booktitle={Proceedings of the 2025 CHI Conference on Human Factors in Computing Systems},
  pages={1--24},
  year={2025}
}

@misc{superrr2024,
  title = {About AI and Unlikelihood},
  author = {{SUPERRR Lab}},
  year = {2024},
  howpublished = {\url{https://superrr.net/en/blog/about-ai-and-unlikelihood}},
  note = {Accessed: 2025-12-01}
}

@misc{manifestno,
  title = {Manifest-No},
  author = {{Manifest-No Collective}},
  year = {2024},
  howpublished = {\url{https://www.manifestno.com/}},
  note = {Accessed: 2025-12-01}
}

@article{dillahunt2025development,
  title={The Development of a New Measure of Collective Digital Literacy},
  author={Dillahunt, Tawanna R and Shedden, Kerby and Filipof, Mila Ekaterina and Lee, Soyoung and Naseem, Mustafa and Toyama, Kentaro and Hui, Julie},
  journal={Proceedings of the ACM on Human-Computer Interaction},
  volume={9},
  number={2},
  pages={1--33},
  year={2025},
  publisher={ACM New York, NY, USA}
}

@article{goyal2022news,
  title={News summarization and evaluation in the era of gpt-3},
  author={Goyal, Tanya and Li, Junyi Jessy and Durrett, Greg},
  journal={arXiv preprint arXiv:2209.12356},
  year={2022}
}

@inproceedings{dillahunt2014fostering,
  title={Fostering social capital in economically distressed communities},
  author={Dillahunt, Tawanna R},
  booktitle={Proceedings of the SIGCHI Conference on Human Factors in Computing Systems},
  pages={531--540},
  year={2014}
}

@article{harrington2019deconstructing,
  title={Deconstructing community-based collaborative design: Towards more equitable participatory design engagements},
  author={Harrington, Christina and Erete, Sheena and Piper, Anne Marie},
  journal={Proceedings of the ACM on human-computer interaction},
  volume={3},
  number={CSCW},
  pages={1--25},
  year={2019},
  publisher={ACM New York, NY, USA}
}

@article{larsson2019independent,
  title={Independent by necessity? The life satisfaction of necessity and opportunity entrepreneurs in 70 countries},
  author={Larsson, Johan P and Thulin, Per},
  journal={Small Business Economics},
  volume={53},
  number={4},
  pages={921--934},
  year={2019},
  publisher={Springer}
}

@inproceedings{delgado2023participatory,
  title={The participatory turn in ai design: Theoretical foundations and the current state of practice},
  author={Delgado, Fernando and Yang, Stephen and Madaio, Michael and Yang, Qian},
  booktitle={Proceedings of the 3rd ACM Conference on Equity and Access in Algorithms, Mechanisms, and Optimization},
  pages={1--23},
  year={2023}
}

@article{mueller2023necessity,
  title={Necessity entrepreneurship: An integrative review and research agenda},
  author={Mueller, Helene and Pieperhoff, Martina},
  journal={Entrepreneurship \& Regional Development},
  volume={35},
  number={9-10},
  pages={762--787},
  year={2023},
  publisher={Taylor \& Francis}
}

@article{hui2018making,
  title={Making a living my way: Necessity-driven entrepreneurship in resource-constrained communities},
  author={Hui, Julie and Toyama, Kentaro and Pal, Joyojeet and Dillahunt, Tawanna},
  journal={Proceedings of the ACM on Human-Computer Interaction},
  volume={2},
  number={CSCW},
  pages={1--24},
  year={2018},
  publisher={ACM New York, NY, USA}
}

@misc{stripe_stripe_nodate,
	title = {Stripe {Atlas}: {Turn} your idea into a startup},
	shorttitle = {Stripe {Atlas}},
	url = {https://stripe.com/atlas},
	language = {en},
	urldate = {2020-03-03},
	author = {Stripe},
	note = {Library Catalog: stripe.com}
}

@article{sum2025you,
  title={" You're in a Ferrari. I'm Waiting for the Bus": Confronting Tensions in Community-University Partnerships},
  author={Sum, Cella M and Zhi, Jiayin and Cook, Amil NT and Cooper, Patrick James and Lozano, Arturo and Johnson, TJ and Perez, Jason and Ghani, Rayid and Skirpan, Michael and Eslami, Motahhare and others},
  journal={Proceedings of the ACM on Human-Computer Interaction},
  volume={9},
  number={2},
  pages={1--28},
  year={2025},
  publisher={ACM New York, NY, USA}
}

@article{glassman2024ai,
  title={Ai-resilient interfaces},
  author={Glassman, Elena L and Gu, Ziwei and Kummerfeld, Jonathan K},
  journal={arXiv preprint arXiv:2405.08447},
  year={2024}
}

@article{benjamin2016informed,
  title={Informed refusal: Toward a justice-based bioethics},
  author={Benjamin, Ruha},
  journal={Science, Technology, \& Human Values},
  volume={41},
  number={6},
  pages={967--990},
  year={2016},
  publisher={SAGE Publications Sage CA: Los Angeles, CA}
}

@article{shail2019using,
  title={Using micro-learning on mobile applications to increase knowledge retention and work performance: a review of literature},
  author={Shail, Mrigank S},
  journal={Cureus},
  volume={11},
  number={8},
  year={2019},
  publisher={Cureus Inc.}
}

@inproceedings{hug2005micro,
  title={Micro Learning and Narration. Exploring possibilities of utilization of narrations and storytelling for the designing of" micro units" and didactical micro-learning arrangements},
  author={Hug, Theo},
  booktitle={fourth Media in Transition conference},
  volume={6},
  number={8},
  year={2005},
  organization={MiT4}
}

@article{otis2023uneven,
  title={The uneven impact of generative AI on entrepreneurial performance},
  author={Otis, Nicholas and Clarke, Rowan P and Delecourt, Solene and Holtz, David and Koning, Rembrand},
  journal={Available at SSRN 4671369},
  year={2023}
}

@inproceedings{kotturi2024deconstructing,
  title={Deconstructing the Veneer of Simplicity: Co-Designing Introductory Generative AI Workshops with Local Entrepreneurs},
  author={Kotturi, Yasmine and Anderson, Angel and Ford, Glenn and Skirpan, Michael and Bigham, Jeffrey P},
  booktitle={Proceedings of the CHI Conference on Human Factors in Computing Systems},
  pages={1--16},
  year={2024}
}

@inproceedings{romero2024exploring,
  title={Exploring the Role of Social Support When Integrating Generative AI in Small Business Workflows},
  author={Romero Lauro, Quentin and Bigham, Jeffrey P and Kotturi, Yasmine},
  booktitle={Companion Publication of the 2024 Conference on Computer-Supported Cooperative Work and Social Computing},
  pages={485--492},
  year={2024}
}

@incollection{riley2005resilience,
  title={Resilience in context},
  author={Riley, Jennifer R and Masten, Ann S},
  booktitle={Resilience in children, families, and communities: Linking context to practice and policy},
  pages={13--25},
  year={2005},
  publisher={Springer}
}

@inproceedings{resnick2008falling,
  title={Falling in love with Seymour’s ideas},
  author={Resnick, Mitchel},
  booktitle={American Educational Research Association (AERA) annual conference},
  year={2008}
}

@misc{openai_chatgpt_overview,
  author       = {{OpenAI}},
  title        = {ChatGPT Overview},
  year         = {2025},
  url          = {https://openai.com/chatgpt/overview/},
  note         = {Accessed: 2025-01-03}
}

@misc{msf365copilot,
    author = {{Microsoft}},
    title={Prompts to try},
    year={2025},
    url={https://support.microsoft.com/en-us/copilot-microsoft365-chat},
    note={Accessed: 2025-04-27}
}

@inproceedings{hui2020community,
  title={Community collectives: Low-tech social support for digitally-engaged entrepreneurship},
  author={Hui, Julie and Barber, Nefer Ra and Casey, Wendy and Cleage, Suzanne and Dolley, Danny C and Worthy, Frances and Toyama, Kentaro and Dillahunt, Tawanna R},
  booktitle={Proceedings of the 2020 CHI conference on human factors in computing systems},
  pages={1--15},
  year={2020}
}

@book{papert2020mindstorms,
  title={Mindstorms: Children, computers, and powerful ideas},
  author={Papert, Seymour A},
  year={2020},
  publisher={Basic books}
}

@misc{sba_business_plan,
  author       = {{U.S. Small Business Administration}},
  title        = {Write Your Business Plan},
  year         = {n.d.},
  url          = {https://www.sba.gov/business-guide/plan-your-business/write-your-business-plan},
  note         = {Accessed: 2025-01-07}
}

@incollection{jackson2002life,
  title={Life in classrooms},
  author={Jackson, Philip},
  booktitle={Teaching and Learning in the Primary School},
  pages={123--128},
  year={2002},
  publisher={Routledge}
}

@inproceedings{zamfirescu2023johnny,
  title={Why Johnny can’t prompt: how non-AI experts try (and fail) to design LLM prompts},
  author={Zamfirescu-Pereira, JD and Wong, Richmond Y and Hartmann, Bjoern and Yang, Qian},
  booktitle={Proceedings of the 2023 CHI Conference on Human Factors in Computing Systems},
  pages={1--21},
  year={2023}
}

@article{bar2007mobile,
  title={Mobile technology appropriation in a distant mirror: Baroque infiltration, creolization and cannibalism},
  author={Bar, Fran{\c{c}}ois and Pisani, Francis and Weber, Matthew},
  journal={Seminario sobre Desarrollo Econ{\'o}mico, Desarrollo Social y Comunicaciones M{\'o}viles en Am{\'e}rica Latina},
  pages={20--21},
  year={2007}
}

@article{hidi2006four,
  title={The four-phase model of interest development},
  author={Hidi, Suzanne and Renninger, K Ann},
  journal={Educational psychologist},
  volume={41},
  number={2},
  pages={111--127},
  year={2006},
  publisher={Taylor \& Francis}
}

@article{avle2019additional,
  title={Additional labors of the entrepreneurial self},
  author={Avle, Seyram and Hui, Julie and Lindtner, Silvia and Dillahunt, Tawanna},
  journal={Proceedings of the ACM on Human-Computer Interaction},
  volume={3},
  number={CSCW},
  pages={1--24},
  year={2019},
  publisher={ACM New York, NY, USA}
}

@online{gusto_2024,
  author       = {Gusto},
  title        = {How SMBs Are Using AI in 2024},
  year         = {2024},
  url          = {https://gusto.com/company-news/smbs-using-ai-2024#},
  urldate      = {2025-01-07},
  note         = {Accessed: 2025-01-07}
}

@online{pewresearch_2023,
  author       = {Pew Research Center},
  title        = {What the Data Says About Americans' Views of Artificial Intelligence},
  year         = {2023},
  url          = {https://www.pewresearch.org/short-reads/2023/11/21/what-the-data-says-about-americans-views-of-artificial-intelligence/},
  urldate      = {2025-01-07},
  note         = {Accessed: 2025-01-07}
}

@inproceedings{kotturi2022tech,
  title={Tech help desk: Support for local entrepreneurs addressing the Long Tail of computing challenges},
  author={Kotturi, Yasmine and Johnson, Herman T and Skirpan, Michael and Fox, Sarah E and Bigham, Jeffrey P and Pavel, Amy},
  booktitle={Proceedings of the 2022 CHI Conference on Human Factors in Computing Systems},
  pages={1--15},
  year={2022}
}

@article{kotturi2024peerdea,
  title={Peerdea: Co-Designing a Peer Support Platform with Creative Entrepreneurs},
  author={Kotturi, Yasmine and Yu, Jenny and Khadpe, Pranav and Gatz, Erin and Zheng, Harvey and Fox, Sarah E and Kulkarni, Chinmay},
  journal={Proceedings of the ACM on Human-Computer Interaction},
  volume={8},
  number={CSCW1},
  pages={1--24},
  year={2024},
  publisher={ACM New York, NY, USA}
}

@article{hui2023community,
  title={Community Tech Workers: Scaffolding Digital Engagement Among Underserved Minority Businesses},
  author={Hui, Julie and Seefeldt, Kristin and Baer, Christie and Sanifu, Lutalo and Jackson, Aaron and Dillahunt, Tawanna R},
  journal={Proceedings of the ACM on Human-Computer Interaction},
  volume={7},
  number={CSCW2},
  pages={1--25},
  year={2023},
  publisher={ACM New York, NY, USA}
}

@inproceedings{dillahunt2018entrepreneurship,
  title={Entrepreneurship and the socio-technical chasm in a lean economy},
  author={Dillahunt, Tawanna R and Kameswaran, Vaishnav and McLain, Desiree and Lester, Minnie and Orr, Delores and Toyama, Kentaro},
  booktitle={Proceedings of the 2018 CHI Conference on Human Factors in Computing Systems},
  pages={1--14},
  year={2018}
}

@inproceedings{abbott2019local,
  title={Local standards for anonymization practices in health, wellness, accessibility, and aging research at CHI},
  author={Abbott, Jacob and MacLeod, Haley and Nurain, Novia and Ekobe, Gustave and Patil, Sameer},
  booktitle={Proceedings of the 2019 CHI Conference on Human Factors in Computing Systems},
  pages={1--14},
  year={2019}
}

@article{gollwitzer1999implementation,
  title={Implementation intentions: strong effects of simple plans.},
  author={Gollwitzer, Peter M},
  journal={American psychologist},
  volume={54},
  number={7},
  pages={493},
  year={1999},
  publisher={American Psychological Association}
}

@article{wong2021reflections,
  title={Reflections on assets-based design: A journey towards a collective of assets-based thinkers},
  author={Wong-Villacres, Marisol and Gautam, Aakash and Tatar, Deborah and DiSalvo, Betsy},
  journal={Proceedings of the ACM on Human-Computer Interaction},
  volume={5},
  number={CSCW2},
  pages={1--32},
  year={2021},
  publisher={ACM New York, NY, USA}
}

@inproceedings{gautam2020crafting,
  title={Crafting, communality, and computing: Building on existing strengths to support a vulnerable population},
  author={Gautam, Aakash and Tatar, Deborah and Harrison, Steve},
  booktitle={Proceedings of the 2020 CHI Conference on Human Factors in Computing Systems},
  pages={1--14},
  year={2020}
}

@inproceedings{pei2019we,
  title={We did it right, but it was still wrong: Toward assets-based design},
  author={Pei, Lucy and Nardi, Bonnie},
  booktitle={Extended Abstracts of the 2019 CHI Conference on Human Factors in Computing Systems},
  pages={1--11},
  year={2019}
}

@article{borchers2025sidehustle,
  author       = {Borchers, Callum},
  title        = {Americans Are Side-Hustling Like We’re in a Recession},
  journal      = {The Wall Street Journal},
  date         = {2025-06-19},
  note         = {Online; accessed 2025-09-07},
}

@online{leppert2024smallbusiness,
  author       = {Leppert, Rebecca},
  title        = {A look at small businesses in the U.S.},
  year         = {2024},
  month        = apr,
  day          = {22},
  publisher    = {Pew Research Center},
  howpublished = {\url{https://www.pewresearch.org/short-reads/2024/04/22/a-look-at-small-businesses-in-the-us/}},
  note         = {Accessed 2025-09-07},
}

@online{liveplan,
  author       = {Iacofano, Chris},
  title        = {LivePlan Assistant},
  organization = {Palo Alto Software},
  year         = {2025},
  url          = {https://help.liveplan.com/hc/en-us/articles/14925894216717-LivePlan-Assistant},
  urldate      = {2025-09-08}
}

@online{upmetrics,
  title        = {AI-Powered Upmetrics Assistant},
  organization = {Upmetrics},
  year         = {2025},
  url          = {https://upmetrics.co/features/ai-assistant},
  urldate      = {2025-09-08}
}

@online{ideabuddy,
  title        = {Smart business plan builder, powered by AI},
  organization = {IdeaBuddy},
  year         = {2025},
  url          = {https://ideabuddy.com/features/business-plan/},
  urldate      = {2025-09-08}
}

@online{score,
  title        = {SCORE: For the Life of Your Business},
  organization = {SCORE Association},
  year         = {2025},
  url          = {https://www.score.org/},
  urldate      = {2025-09-08}
}

@article{masten2001ordinary,
  title={Ordinary magic: Resilience processes in development.},
  author={Masten, Ann S},
  journal={American psychologist},
  volume={56},
  number={3},
  pages={227},
  year={2001},
  publisher={American Psychological Association}
}

@article{vyas2017everyday,
  title={Everyday resilience: Supporting resilient strategies among low socioeconomic status communities},
  author={Vyas, Dhaval and Dillahunt, Tawanna},
  journal={Proceedings of the ACM on Human-Computer Interaction},
  volume={1},
  number={CSCW},
  pages={1--21},
  year={2017},
  publisher={ACM New York, NY, USA}
}

@article{muller2002participatory,
author = {Muller, Michael and Druin, Allison},
year = {2002},
month = {01},
pages = {},
title = {Participatory Design: The Third Space in HCI},
journal = {Handbook of HCI}
}

@inproceedings{gautam2024reconfiguring,
  title={Reconfiguring participatory design to resist ai realism},
  author={Gautam, Aakash},
  booktitle={Proceedings of the Participatory Design Conference 2024: Exploratory Papers and Workshops-Volume 2},
  pages={31--36},
  year={2024}
}

@inproceedings{birhane2022power,
  title={Power to the people? Opportunities and challenges for participatory AI},
  author={Birhane, Abeba and Isaac, William and Prabhakaran, Vinodkumar and Diaz, Mark and Elish, Madeleine Clare and Gabriel, Iason and Mohamed, Shakir},
  booktitle={Proceedings of the 2nd ACM Conference on Equity and Access in Algorithms, Mechanisms, and Optimization},
  pages={1--8},
  year={2022}
}

@inproceedings{cooper2022systematic,
  title={A systematic review and thematic analysis of community-collaborative approaches to computing research},
  author={Cooper, Ned and Horne, Tiffanie and Hayes, Gillian R and Heldreth, Courtney and Lahav, Michal and Holbrook, Jess and Wilcox, Lauren},
  booktitle={Proceedings of the 2022 CHI conference on human factors in computing systems},
  pages={1--18},
  year={2022}
}

@article{liang2021embracing,
  title={Embracing four tensions in human-computer interaction research with marginalized people},
  author={Liang, Calvin A and Munson, Sean A and Kientz, Julie A},
  journal={ACM Transactions on Computer-Human Interaction (TOCHI)},
  volume={28},
  number={2},
  pages={1--47},
  year={2021},
  publisher={ACM New York, NY, USA}
}

@article{kapur2008productive,
  title={Productive failure},
  author={Kapur, Manu},
  journal={Cognition and instruction},
  volume={26},
  number={3},
  pages={379--424},
  year={2008},
  publisher={Taylor \& Francis}
}

@article{murdoch2020feeling,
  title={Feeling heard: Inclusive education, transformative learning, and productive struggle},
  author={Murdoch, Diana and English, Andrea R and Hintz, Allison and Tyson, Kersti},
  journal={Educational theory},
  volume={70},
  number={5},
  pages={653--679},
  year={2020},
  publisher={Wiley Online Library}
}

@inproceedings{inman2019beautiful,
  title={" Beautiful Seams" Strategic Revelations and Concealments},
  author={Inman, Sarah and Ribes, David},
  booktitle={Proceedings of the 2019 CHI Conference on Human Factors in Computing Systems},
  pages={1--14},
  year={2019}
}

@article{warshauer2015productive,
  title={Productive struggle in middle school mathematics classrooms},
  author={Warshauer, Hiroko Kawaguchi},
  journal={Journal of Mathematics Teacher Education},
  volume={18},
  number={4},
  pages={375--400},
  year={2015},
  publisher={Springer}
}

@inproceedings{haliburton2024longitudinal,
  title={A longitudinal in-the-wild investigation of design frictions to prevent smartphone overuse},
  author={Haliburton, Luke and Gr{\"u}ning, David Joachim and Riedel, Frederik and Schmidt, Albrecht and Terzimehi{\'c}, Na{\dj}a},
  booktitle={Proceedings of the 2024 CHI Conference on Human Factors in Computing Systems},
  pages={1--16},
  year={2024}
}

@inproceedings{gautam2020p,
  title={p for political: Participation without agency is not enough},
  author={Gautam, Aakash and Tatar, Deborah},
  booktitle={Proceedings of the 16th Participatory Design Conference 2020-Participation (s) Otherwise-Volume 2},
  pages={45--49},
  year={2020}
}

@inproceedings{vines2013configuring,
  title={Configuring participation: on how we involve people in design},
  author={Vines, John and Clarke, Rachel and Wright, Peter and McCarthy, John and Olivier, Patrick},
  booktitle={Proceedings of the SIGCHI conference on human factors in computing systems},
  pages={429--438},
  year={2013}
}

@article{arnstein1969ladder,
  title={A ladder of citizen participation},
  author={Arnstein, Sherry R},
  journal={Journal of the American Institute of planners},
  volume={35},
  number={4},
  pages={216--224},
  year={1969},
  publisher={Taylor \& Francis}
}

@misc{CloudBank,
  title        = {CloudBank: Cloud Computing for Research \& Education},
  howpublished = {\url{https://www.cloudbank.org/}},
  note         = {Accessed: 2025-09-11},
  institution  = {CloudBank},
  year         = {2025}
}

@misc{OpenRouter,
  title        = {OpenRouter: The Unified Interface for LLMs},
  howpublished = {\url{https://openrouter.ai}},
  note         = {Accessed: 2025-09-11},
  institution  = {OpenRouter, Inc.},
  year         = {2025}
}

@misc{openai-openapi,
  author       = {OpenAI},
  title        = {openai/openai-openapi: OpenAPI specification for the OpenAI API},
  howpublished = {\url{https://github.com/openai/openai-openapi}},
  note         = {Accessed: 2025-09-11},
  year         = {2025},
  license      = {MIT}
}

@inproceedings{DillahuntEtAl_CHI22_VillageMentoring,
  author = {Dillahunt, Tawanna R. and Lu, Alex Jiahong and Israni, Aarti and Lodha, Ruchita and Brewer, Savana and Robinson, Tiera S. and Wilson, Angela Brown and Wheeler, Earnest},
  title = {The Village: Infrastructuring Community-based Mentoring to Support Adults Experiencing Poverty},
  booktitle = {CHI Conference on Human Factors in Computing Systems (CHI ’22)},
  year = {2022},
  pages = {1--17},  
  publisher = {Association for Computing Machinery},
  address = {New Orleans, LA, USA},
  doi = {10.1145/3491102.3501949}
}

@article{BUCKNELL2022111234,
title = {Adaptive self-reflection and resilience: The moderating effects of rumination on insight as a mediator},
journal = {Personality and Individual Differences},
volume = {185},
pages = {111234},
year = {2022},
issn = {0191-8869},
doi = {https://doi.org/10.1016/j.paid.2021.111234},
url = {https://www.sciencedirect.com/science/article/pii/S0191886921006139},
author = {K.J. Bucknell and M. Kangas and M.F. Crane}
}

@article{Cook2021RiseOfTikTokEntrepreneurs,
  author       = {Shannon Cook},
  title        = {The Rise Of TikTok Entrepreneurs: How Useful Is Their Business Advice?},
  journal      = {BusinessBecause},
  year         = {2021},
  month        = {May},
  day          = {27},
  url          = {https://www.businessbecause.com/news/insights/7658/tiktok-entrepreneurs-business-advice},
  note         = {Accessed: 2025-09-11}
}

@book{vonhippel2006democratizing,
  author    = {Eric von Hippel},
  title     = {Democratizing Innovation},
  year      = {2006},
  publisher = {The MIT Press},
  address   = {Cambridge, MA}
}

@article{Li02092025,
author = {Keyao Li and Mark A. Griffin and Mengting (Rachel) Xia},
title = {How do workforce adaptability and reskilling initiatives drive innovations: the case of Western Australian construction industry},
journal = {Construction Management and Economics},
volume = {43},
number = {9},
pages = {746--763},
year = {2025},
publisher = {Routledge},
doi = {10.1080/01446193.2025.2511831},
URL = {https://doi.org/10.1080/01446193.2025.2511831},
eprint = {https://doi.org/10.1080/01446193.2025.2511831}
}

@article{Bonanno2004LossTrauma,
  author  = {George A. Bonanno},
  title   = {Loss, Trauma, and Human Resilience: Have We Underestimated the Human Capacity to Thrive After Extremely Aversive Events?},
  journal = {American Psychologist},
  year    = {2004},
  volume  = {59},
  number  = {1},
  pages   = {20--28},
  doi     = {10.1037/0003-066X.59.1.20}
}

@article{Rutter2012ResilienceDynamic,
  author  = {Michael Rutter},
  title   = {Resilience as a Dynamic Concept},
  journal = {Development and Psychopathology},
  year    = {2012},
  volume  = {24},
  number  = {2},
  pages   = {335--344},
  doi     = {10.1017/S0954579412000028}
}

@article{Luthar1991VulnerabilityResilience,
  author  = {Suniya S. Luthar},
  title   = {Vulnerability and Resilience: A Study of High-Risk Adolescents},
  journal = {Child Development},
  year    = {1991},
  volume  = {62},
  number  = {3},
  pages   = {600--616},
  doi     = {10.1111/j.1467-8624.1991.tb01555.x}
}

@article{Ungar2013ResilienceContext,
  author  = {Michael Ungar},
  title   = {Resilience, Trauma, Context, and Culture},
  journal = {Trauma, Violence, \& Abuse},
  year    = {2013},
  volume  = {14},
  number  = {3},
  pages   = {255--266},
  doi     = {10.1177/1524838013487805}
}

@article{Southwick2014ResilienceDefinitions,
  author  = {Steven M. Southwick and George A. Bonanno and Ann S. Masten and Catherine Panter-Brick and Rachel Yehuda},
  title   = {Resilience Definitions, Theory, and Challenges: Interdisciplinary Perspectives},
  journal = {European Journal of Psychotraumatology},
  year    = {2014},
  volume  = {5},
  pages   = {25338},
  doi     = {10.3402/ejpt.v5.25338}
}

@article{Patel2017CommunityResilience,
  author  = {Shantanu S. Patel and Monica S. Rogers and Jason Amlôt and G. James Rubin},
  title   = {What Do We Mean by “Community Resilience”? A Systematic Literature Review of How It Is Defined in the Literature},
  journal = {PLoS Currents},
  year    = {2017},
  volume  = {9},
  doi     = {10.1371/currents.dis.db775aff25efc5ac4f0660ad9c9f7db2}
}

@article{Norris2007CommunityResilience,
  author  = {Fran H. Norris and Susan P. Stevens and Betty Pfefferbaum and Karen F. Wyche and Rose L. Pfefferbaum},
  title   = {Community Resilience as a Metaphor, Theory, Set of Capacities, and Strategy for Disaster Readiness},
  journal = {American Journal of Community Psychology},
  year    = {2007},
  volume  = {41},
  pages   = {127--150},
  doi     = {10.1007/s10464-007-9156-6}
}

@article{Israel1998Review,
  author    = {Barbara A. Israel and Amy J. Schulz and Edith A. Parker and Adam B. Becker},
  title     = {Review of Community-Based Research: Assessing Partnership Approaches to Improve Public Health},
  journal   = {Annual Review of Public Health},
  year      = {1998},
  volume    = {19},
  pages     = {173--202},
  doi       = {10.1146/annurev.publhealth.19.1.173}
}

@article{Morris2023DesignSpaceGenerativeModels,
  author    = {Meredith Ringel Morris and Carrie J. Cai and Jess Holbrook and Chinmay Kulkarni and Michael Terry},
  title     = {The Design Space of Generative Models},
  journal   = {arXiv preprint arXiv:2304.10547},
  year      = {2023},
  url       = {https://arxiv.org/abs/2304.10547},
  note      = {Accessed: 2025-09-11}
}

@article{KuhnGalloway2015PeerNetworking,
  author    = {Kristine M. Kuhn and Tera L. Galloway},
  title     = {With a Little Help from My Competitors: Peer Networking among Artisan Entrepreneurs},
  journal   = {Entrepreneurship Theory and Practice},
  year      = {2015},
  volume    = {39},
  number    = {3},
  pages     = {571--600},
  doi       = {10.1111/etap.12053}
}

@article{Kay2023MakerspacesEffectiveLearning,
  author    = {Louise Kay and Amy Buxton},
  title     = {Makerspaces and the Characteristics of Effective Learning in the Early Years},
  journal   = {Journal of Early Childhood Research},
  year      = {2023},
  volume    = {22},
  number    = {3},
  pages     = {343--358},
  doi       = {10.1177/1476718X231210633},
  note      = {Original work published 2024}
}

@book{Pirolli2007InformationForaging,
  author    = {Peter Pirolli},
  title     = {Information Foraging Theory: Adaptive Interaction with Information},
  year      = {2007},
  publisher = {Oxford University Press},
  address   = {Oxford, UK}
}

@misc{SBA2025WriteBusinessPlan,
  author       = {U.S. Small Business Administration},
  title        = {Write Your Business Plan: Sample Business Plans \& Business Plan Template},
  year         = {2025},
  howpublished = {\url{https://www.sba.gov/business-guide/plan-your-business/write-your-business-plan}},
  note         = {Accessed: 2025-09-11}
}

@article{Flanagan1954CriticalIncident,
  author  = {John C. Flanagan},
  title   = {The Critical Incident Technique},
  journal = {Psychological Bulletin},
  year    = {1954},
  volume  = {51},
  number  = {4},
  pages   = {327--358},
  doi     = {10.1037/h0061470}
}

@article{bhargava2015beyond,
  title={Beyond data literacy: Reinventing community engagement and empowerment in the age of data},
  author={Bhargava, Rahul and Deahl, Erica and Letouz{\'e}, Emmanuel and Noonan, Amanda and Sangokoya, David and Shoup, Natalie},
  year={2015},
  publisher={Data Pop Alliance}
}

@inproceedings{long2020ai,
  title={What is AI literacy? Competencies and design considerations},
  author={Long, Duri and Magerko, Brian},
  booktitle={Proceedings of the 2020 CHI conference on human factors in computing systems},
  pages={1--16},
  year={2020}
}

@incollection{baumer2014refusing,
  title={Refusing, limiting, departing: why we should study technology non-use},
  author={Baumer, Eric PS and Ames, Morgan G and Brubaker, Jed R and Burrell, Jenna and Dourish, Paul},
  booktitle={CHI'14 Extended Abstracts on Human Factors in Computing Systems},
  pages={65--68},
  year={2014}
}

@inproceedings{satchell2009beyond,
  title={Beyond the user: use and non-use in HCI},
  author={Satchell, Christine and Dourish, Paul},
  booktitle={Proceedings of the 21st annual conference of the Australian computer-human interaction special interest group: Design: Open 24/7},
  pages={9--16},
  year={2009}
}

@article{selwyn2006digital,
  title={Digital division or digital decision? A study of non-users and low-users of computers},
  author={Selwyn, Neil},
  journal={Poetics},
  volume={34},
  number={4-5},
  pages={273--292},
  year={2006},
  publisher={Elsevier}
}

@inproceedings{gautam2022empowering,
  title={Empowering Participation Within Structures of Dependency},
  author={Gautam, Aakash and Tatar, Deborah},
  booktitle={Proceedings of the Participatory Design Conference 2022-Volume 1},
  pages={75--86},
  year={2022}
}

@inproceedings{pierre2021getting,
  title={Getting Ourselves Together: Data-centered participatory design research \& epistemic burden},
  author={Pierre, Jennifer and Crooks, Roderic and Currie, Morgan and Paris, Britt and Pasquetto, Irene},
  booktitle={Proceedings of the 2021 CHI Conference on Human Factors in Computing Systems},
  pages={1--11},
  year={2021}
}

\end{document}